\documentclass[preprint,showpacs,titlepage,aps,prd,
tightenlines,byrevtex,nofootinbib]{revtex4}

\usepackage{graphicx,amssymb}

\def\GWth{\text{GW}_{\text{th}}}

\def\J-PARC{{\text{J-PARC}}}

\def\CP1{{\text{CP1}}}

\def\lsim{\raise0.3ex\hbox{$\;<$\kern-0.75em\raise-1.1ex
\hbox{$\sim\;$}}}
\def\gsim{\raise0.3ex\hbox{$\;>$\kern-0.75em\raise-1.1ex
\hbox{$\sim\;$}}}
\begin{document}

\preprint{hep-ph/0407326}
\title{
Reactor Measurement of $\theta_{12}$; \\
Principles, Accuracies and Physics Potentials}

\author{H. Minakata$^{1}$}
\email{minakata@phys.metro-u.ac.jp}
\author{H. Nunokawa$^{2}$}
\email{nunokawa@fis.puc-rio.br}
\author{W. J. C. Teves$^3$}\email{teves@fma.if.usp.br} 
\author{R. Zukanovich Funchal$^{3}$}\email{zukanov@if.usp.br} 
\affiliation{
$^1$Department of Physics, Tokyo Metropolitan University, 
Hachioji, Tokyo 192-0397, Japan \\
$^2$Pontif{\'\i}cia Universidade Cat{\'o}lica do Rio de Janeiro, \\
C. P. 38071, 22452-970, Rio de Janeiro, Brazil \\
$^3$Instituto de F{\'\i}sica,  Universidade de S{\~a}o Paulo, 
C.\ P.\ 66.318, 05315-970, S{\~a}o Paulo, Brazil 
}

\date{December 31, 2004}

\vglue 1.4cm

\begin{abstract}

We discuss reactor measurement of $\theta_{12}$ which has a potential
of reaching the ultimate sensitivity which surpasses all the methods
so far proposed. The key is to place a detector at an appropriate
baseline distance from the reactor neutrino source to have an
oscillation maximum at around a peak energy of the event spectrum in
the absence of oscillation.  By a detailed statistical analysis the
optimal distance is estimated to be 
$\simeq (50-70)$ km$\times 
[8 \times 10^{-5} \text{eV}^2/\Delta m^2_{21}]$, 
which is determined by maximizing the oscillation effect in the event 
number distribution and minimizing geo-neutrino background contamination. 
To estimate possible uncertainty caused by surrounding nuclear reactors in
distance of $\sim 100$ km, we examine a concrete example of a detector
located at Mt. Komagatake, 54 km away from the Kashiwazaki-Kariwa
nuclear power plant in Japan, the most powerful reactor complex in the
world. The effect turns out to be small. 
Under a reasonable assumption of systematic error of 4\% in the experiment, 
we find that $\sin^2{\theta_{12}}$ can be determined to the accuracy of 
$\simeq $ 2\% ($\simeq$ 3\%), at 68.27\% CL for 1 degree of freedom, for 60
GW$_{\text{th}}\cdot$kton$\cdot$yr (20 GW$_{\text{th}}\cdot$kton$\cdot$yr)
operation. We also discuss implications of such an accurate
measurement of $\theta_{12}$.

\end{abstract}


\maketitle


\section{Introduction}

Reactor neutrino experiments have been the source of rich 
physics informations \cite{review1}. The very existence of 
(anti-)neutrino was proven by the memorial reactor neutrino 
experiment by Reines and Cowan \cite{reines-cowan}.
In recent years, reactor neutrino experiments have been playing 
greater r{\^o}le than in any other era of neutrino physics. 
Following SNO \cite{SNO,sno_salt} who has confirmed by the 
{\it in situ} measurement that the solar neutrino deficit is 
indeed caused by neutrino flavor transformation, 
KamLAND \cite{KamLAND,KamLAND_new}  observed clear 
deficit of reactor neutrinos 
and pinned down a unique parameter region, the large-mixing-angle 
(LMA) Mikheyev-Smirnov-Wolfenstein (MSW) solution \cite{MSW}.
Thus, the solar neutrino problem which lasted nearly 40 years 
has now been solved. At the same time, the basic structure 
of the (1-2) sector of the lepton flavor mixing matrix, 
the Maki-Nakagawa-Sakata (MNS) matrix \cite{MNS}, was determined.

Reactor neutrino experiments also play (and will play) crucial 
r{\^o}le to explore the (1-3) sector of the MNS matrix, the unique 
sector whose structure is not yet determined. 
The CHOOZ and the Palo Verde experiments were able to place 
stringent constraints on $\theta_{13}$ \cite{CHOOZ,PaloVerde}. 
It was recognized that reactor measurement of $\theta_{13}$ 
has very special characteristics as a pure measurement of 
$\theta_{13}$, whose property should play an important part  
in solving the problem of parameter degeneracy \cite{MSYIS}. 
It was also stressed that the r{\^o}le played by reactor experiments 
is complementary to those by long-baseline (LBL) accelerator 
neutrino experiments. 
The proposal, sometime after an earlier Russian project~\cite{krasnoyarsk}, 
was followed by a spur of experimental projects over the globe which 
are now summarized in the White Paper Report \cite{reactor_white}.

Are those described above all that can be done by reactor 
neutrino experiments? We answer the question in the negative 
by proposing dedicated reactor neutrino experiments for precise 
measurement of $\theta_{12}$ in this paper. 
We will show that the accuracy attainable by such reactor experiments 
for $\theta_{12}$ can reach to $\simeq$ 2\% for $\sin^2\theta_{12}$
for 1 degree of freedom (DOF)
under a reasonable assumption of systematic errors. 
It surpasses those expected by the other methods 
so far proposed, e.g., by combining KamLAND with accurate 
$^7$Be and pp solar neutrino measurement \cite{bahcall-pena}.
Throughout this paper, we will demonstrate these statements by careful 
treatment of the optimization of the baseline distance as well as the effects 
of geophysical neutrinos (hereafter, geo-neutrinos).

The key idea for such an enormous sensitivity is to place the detector 
at the appropriate distance to see the effect of oscillation at the place 
where it is maximal. A zeroth order estimation of the distance 
\cite{BCG03} is given by 
$L_{\text{OM}} \equiv 2\pi E_\nu^{\text{peak}}/\Delta m^2_{21} \simeq 63 $ 
km, where $\Delta m^2_{21} = 7.9 \times 10^{-5} \mbox{eV}^2$ is 
the current best fit value \cite{KamLAND_new}, and 
$E_\nu^{\text{peak}}=4$ MeV is a typical neutrino energy where 
event rate has a maximum in the absence of oscillation. 
But, due to the fact that the reactor neutrino energy spectrum is 
rather broad and the $1/L^2$ dependence of the intensity of neutrino flux, 
the optimal distances turned out to be shorter and spread over a range of 
50 to 70 km, adding more variety for the site selection of the detector. 
We also discuss the geo-neutrino background contamination and its 
relevance in determining the best position to place the detector.

It seems that now a coherent view of how to determine accurately 
the ``large" mixing angles has emerged. 
Namely, disappearance measurement at around the oscillation maximum 
gives the highest sensitivities for both of the large angles, 
$\theta_{12}$ and $\theta_{23}$. 
It is well recognized that the most accurate way of determining 
$\sin^22\theta_{23}$ will be achieved by the next generation LBL
accelerator neutrino experiments by using their $\nu_{\mu}$ 
disappearance mode \cite{JPARC,NuMI,SPL}. 
In particular, in the JPARC-SK experiment, the accuracy of
determination of $\sin^2{2\theta_{23}}$ is expected to reach
down to $\simeq$ 1\% at 90\% CL \cite{JPARC,MSS}.

What is the scientific merit of such precise measurement of $\theta_{12}$?
We will discuss this question in length in Sec.~\ref{implication}, 
but here we make only two remarks, one from particle physics 
and the other from solar physics point of view. 
Current understanding of nonzero neutrino mass involves, most probably, 
the existence of a large energy scale where leptons and quarks are unified 
\cite{GUT}, whose simplest model realization is the seesaw model \cite{seesaw}.
Various models so far proposed try to relate quark and lepton mass matrix 
based on the underlying philosophy. For a review see e.g., \cite{altarelli}.
We will probably need detailed informations of lepton mixing matrix 
which are comparable with that of quarks when we want to test the theory for lepton 
and quark mixings, which should emerge someday in the future.
If the solar angle $\theta_{12}$ is in fact complementary to the 
Cabibbo angle \cite{QLC}, there is an immense request for accurate 
determination of $\theta_{12}$.

What is the solar physics implication of accurate measurement of $\theta_{12}$?
The solar neutrino flux measured on the earth inevitably contains the effect
of neutrino flavor transformation, and therefore the precision of its 
determination is affected by the uncertainty of $\theta_{12}$. 
We thus believe that accurate determination of $\theta_{12}$ has a great merit 
for precise observational solar astrophysics in which an accurate measurement 
of the infant flux produced at the solar core is required.

In Sec.~\ref{principle} we discuss the basic principle and power of the method of tuning 
the baseline distance for accurate determination of $\theta_{12}$ by reactor 
neutrinos. 
In Sec.~\ref{requirement} we discuss the requirements and possible 
obstacle for such reactor $\theta_{12}$ experiments. 
They include the systematic errors, a possible uncertainty due to 
geo-neutrino background and the unknown value of $\theta_{13}$.
In Sec.~\ref{procedure} we describe the analysis procedure and 
give a quantitative estimate of the optimal baseline distances by 
fully taking into account the geo-neutrino background.
We use a specific setting of the detector to estimate effects of the other 
surrounding reactors. 
In Sec.~\ref{analysis} we carry out the sensitivity analysis.
In Sec.~\ref{stability} we examine the stability of our results 
against changes of our statistical procedure. 
In Sec.~\ref{comparison} we compare the sensitivity of reactor 
$\theta_{12}$ experiments with the one that will be reached by 
KamLAND-solar combined method. 
In Sec.~\ref{implication} we discuss possible physics implications of a  
precise measurement of $\theta_{12}$.

\section{Accurate Measurement of $\theta_{12}$ by Reactors by 
tuning baseline distance}
\label{principle}

We discuss in this section the method for accurate determination of 
$\theta_{12}$ by choosing an appropriate distance to the detector from a 
principal (nearest) nuclear reactor. 
We start by giving a pedagogical self-contained description of the basic 
principle of the measurement. 
We then contrast the sensitivity achievable by our method 
with that of KamLAND \cite{KamLAND,KamLAND_new} to illuminate the 
power of the method we propose by explicitly showing improvement 
over the marvelous experiment using the {\it same} neutrino flux from 
the reactors located all over Japan.

For ease in frequent reference to a dedicated liquid scintillator detector at the 
distances of 50-70 km, we use in this paper the acronym ``SADO'' 
as an abbreviation,\footnote{
  ``SADO" was originally created as an acronym for a detector in Sado
  island in Niigata, Japan \cite{mina_niigata}, which is located at about 
  71 km from the Kashiwazaki-Kariwa nuclear reactor complex, 
  whose distance roughly corresponds to the oscillation maximum for 
  the old best fit value of
  $\Delta m^2_{21} = 7.2 \times 10^{-5}$ eV$^2$ \cite{KamLAND}.
}
\begin{equation}
\mbox{SADO} \equiv \mbox{Several-tens of km Antineutrino DetectOr}.
\end{equation}

\subsection{Method for precise measurement of $\theta_{12}$}  

The secret behind the potentially powerful method is to 
place a detector to the location where 
the oscillation effect has its maximum in the observable 
quantity, i.e., the number of events (not in the probability alone). 
The optimal distances for a given value of $\Delta m^2_{21}$ are
\begin{equation}
L_{\rm optimal} \simeq (50-70) 
\left[\displaystyle \frac{\Delta m^2_{21}}
{8 \times 10^{-5}\ \text{eV}^2} \right]^{-1} \text{km,}
\label{eq:scale}
\end{equation}
as we will find in Sec.~\ref{procedure} after establishing our 
statistical method (See Fig.~\ref{dist-dep}.) 
The optimal distance coincides approximately, but not exactly,  
with the first oscillation maximum 
(the one encountered first by neutrinos as they travel)
for the peak energy of reactor neutrino event distribution. 
Moreover, such a distance must be established by 
taking into account background contribution 
from geo-neutrinos, as we will see in Sec.~\ref{requirement}. 

\begin{figure}[htbp]
\vglue -0.5cm
\begin{center}
\includegraphics[width=0.65\textwidth]{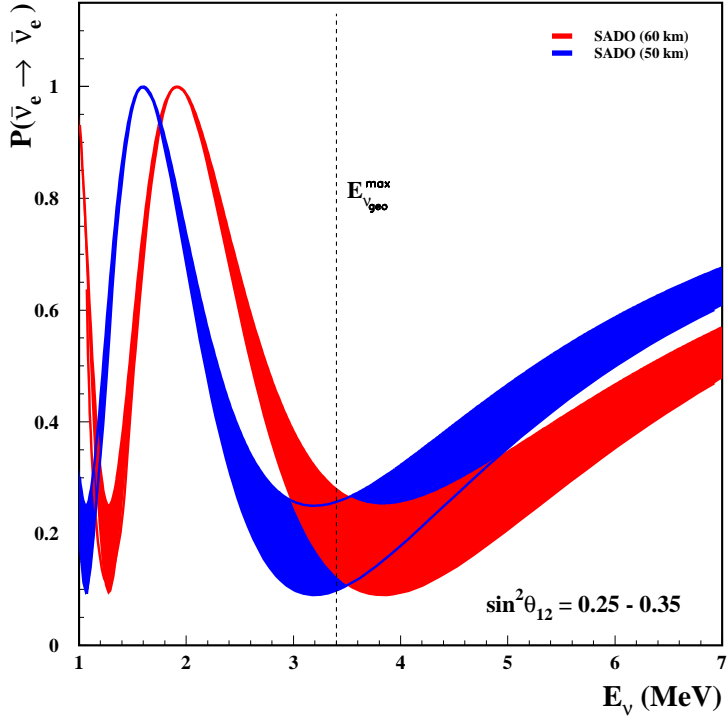}
\end{center}
\vglue -0.7cm
\caption{
Ranges of the survival probabilities 
$P(\bar{\nu}_{e} \rightarrow \bar{\nu}_{e})$ 
of antineutrinos from the nuclear reactor complex 
as a function of neutrino energy
are plotted for two distances to SADO, 50 km (blue) and 60 km (red). 
The upper and lower boundary of each band 
corresponds, respectively, to $\sin^2 \theta_{12} = 0.25$ 
and 0.35.  
Here $\Delta m^2_{21}=7.9\times 10^{-5}$ eV$^2$ and $\theta_{13}=0$. 
The vertical dashed line marks the current KamLAND energy threshold 
$E_{\nu}=3.4$ MeV for reactor neutrino analysis, which corresponds to 
the maximum energy for geo-neutrinos. 
}
\label{prob_kam-sado}
\end{figure}

In Fig.\ref{prob_kam-sado} we show the range of survival probabilities 
as a function of neutrino energy in a form of band which is spanned by 
$P(\bar{\nu}_{e} \rightarrow \bar{\nu}_{e})$ obtained by varying 
$\sin^2 \theta_{12}$ between 0.25 (upper end) and 0.35 (lower end). 
The probabilities are computed 
using the current best fit value of $\Delta m^2_{21}$ and 
with the two distances $L= 50$ (red) and 60 km (blue).

We observe that the oscillation maximum (minimum in the survival probability) 
occurs at $E_\nu \simeq (3.0$-4.0) MeV around the peak energy where 
the event rate is maximal in the absence of oscillation. 
The most important feature for us is that the depth of the minimum in 
the survival probability is the most sensitive 
place to the variation of $\sin^2\theta_{12}$ in the entire figure. 
Nothing but that property is the key to the enormous accuracy 
of reactor $\theta_{12}$ measurement to be explored in this paper.

We should note that there were some previous attempts 
along the similar line of thought.
To our knowledge, dedicated reactor experiment for $\theta_{12}$ at 
the oscillation maximum has first been considered by Takasaki 
who did a back of the envelope estimation of the sensitivity assuming 
an appropriate site at $\sim$70 km away from the Fukushima nuclear 
reactors \cite{takasaki}.
A statistical treatment for estimating the accuracy of $\theta_{12}$ 
determination was presented by Bandyopadhyay {\it et al.}~\cite{BCG03} 
which resulted in  about 10\% error (99\% CL, 2 DOF) in 
$\sin^2\theta_{12}$. It is about factor of 2 larger than ours; 
See Fig.~\ref{sado54a}.
They also did not address the optimization problem of the sensitivity 
with respect to the baseline distance. 
A different strategy which utilizes not only the first oscillation maximum 
but also the subsequent minimum was proposed by Bouchiat 
\cite{bouchiat}.

\subsection{Setting}

Throughout this paper we analyze simultaneously the two settings: 

\vskip 0.2cm

\noindent
(A) SADO with single reactor complex, which will be denoted as 
SADO$_{\rm single}$. 

\vskip 0.2cm

\noindent
(B) SADO with multiple reactor complex, which will be denoted as
SADO$_{\rm multi}$. 

\vskip 0.2cm

\noindent
The set up (A) may be an excellent approximation for 
many reactor sites in the world including Angra reactor 
in Brazil and Daya Bay in China~\cite{reactor_white}. 
On the other hand, the set up (B) with multiple reactor 
complex in addition to the nearest one, which could act 
as ``background", is inevitable if we 
think about experiments in Japan, for example, where many nuclear 
power plants (NPPs) are located within $\sim$ 100 km.

For definiteness, we take a particular site for the set up (B) 
in this paper; SADO at Mt. Komagatake, Niigata prefecture, Japan, 
which is about 54 km away from the Kashiwazaki-Kariwa nuclear 
reactor complex with the maximal thermal power of 24.3 GW$_{\rm th}$, 
the largest NPP in the world.\footnote{
The site at $\sim 1000$ m below the peak of Mt. Komagatake is 
tentatively selected because of the appropriate distance from the 
Kashiwazaki-Kariwa NPP, 
a sufficient overburden, and possibility of relatively short access of 
about 3-5 km from a public road by digging a tunnel. 
Nonetheless, the site is not meant to be unique and there exist many 
other potential sites in the strip of distance 50-70 km from the reactor 
because the region is quite mountainous. 
}
Taking the particular setting is necessary to estimate the 
``background effects'' caused by the other reactors than 
the nearest one. 
For the former case (A) one can use any reactor site, but for 
purpose of comparison, we just switch off contributions 
from the other 15 reactors except for the Kashiwazaki-Kariwa 
NPP to use the common normalization factor.
In Table~\ref{Tabzero} in Appendix~\ref{ap1}, we present the names 
of the reactor NPPs, their 
current (near future) thermal power, their distances from KamLAND and 
from SADO as well as their relative contributions to the neutrino flux
at these sites.

To make our analysis useful for experiments at any other reactor 
sites we present our plots, except for Fig.\ref{kl-sado-90-kashi}, 
in units of GW$_{\rm th}\cdot$kt$\cdot$yr 
where ``kt'' is an abbreviation of kton.
Notice that the number in the unit refers only to the reactor power 
of the principal reactor, the Kashiwazaki-Kariwa NPP in this paper, and 
does not include the one from other 15 reactors.
In this work, we ignore all the Japanese research reactors 
as well as any other reactors outside Japan. In fact, 
their contribution to the $\bar{\nu}_e$ flux at KamLAND  is 
calculated to be 4.5\%~\cite{KamLAND_new}, while the corresponding 
value at SADO is significantly smaller, about 1.1\%
due to the much larger flux from the closest reactor. 
We do not take into account these contributions explicitly 
in our calculation assuming that they can be subtracted from data. 
We can argue, quite reasonably, that the uncertainty of estimation 
of such contribution is at most 10-20\%, and the additional 
systematic error due to the subtraction is negligibly small.

\subsection{KamLAND vs. SADO}

One of the crucial questions for us is ``to what extent is 
the principle of tuning the experiment to 
the oscillation maximum effective in 
improving the accuracy of determination of $\theta_{12}$?''
The answer to this question will also tell us 
if SADO can supersede KamLAND, and if yes, to what extent.

\begin{figure}[htbp]
\vglue -3cm
\begin{center}
\includegraphics[width=1.2\textwidth]{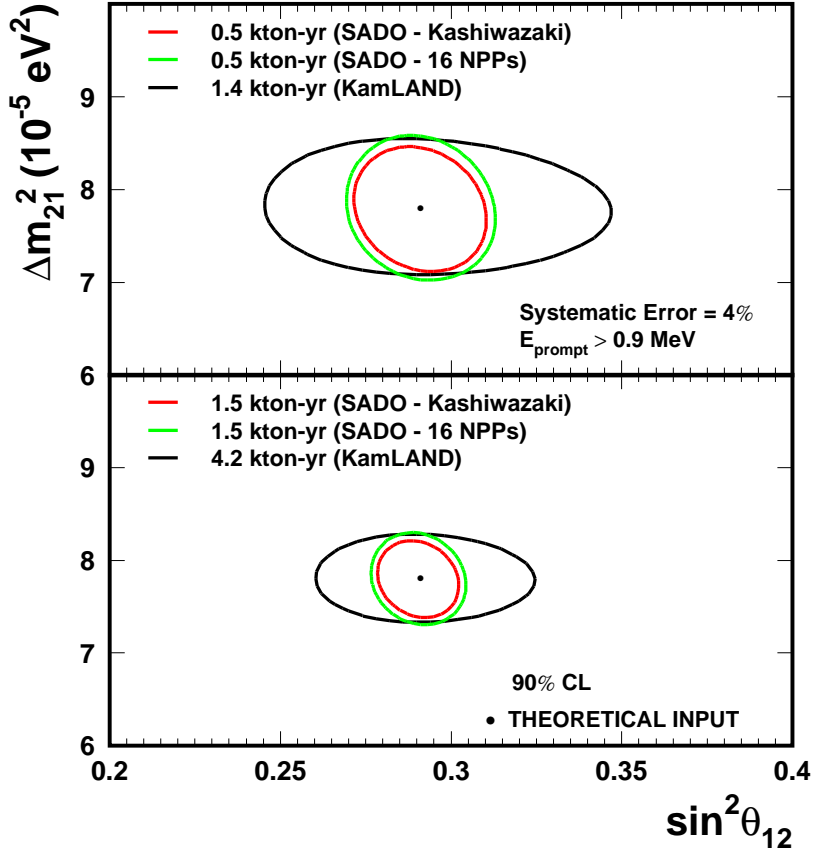}
\end{center}
\vglue -5.5cm
\caption{
Comparison between the sensitivities at 90\% CL  
for 2 DOF expected by KamLAND and SADO 
for runs with (about) equal number of 
events without oscillation.  Neutrinos from
all the 16 NPPs are taken into account for KamLAND
whereas for SADO$_{\rm single}$ only Kashiwazaki-Kariwa NPP contribution was 
computed.  The input values of mixing parameters are
$\sin^2 \theta_{12}= 0.29$ and 
$\Delta m^2_{21} = 7.9 \times 10^{-5}$ eV$^2$. 
The numbers of events for 0.5 kt$\cdot$yr 
measurement at SADO$_{\rm single}$ and for 
1.4 kt$\cdot$yr measurement at KamLAND will be about 1700. 
The systematic errors are assumed to be 4\% in both detectors.  
Events with prompt energy 
($E_{\text{prompt}} \equiv E_{\nu}$ - 0.8 MeV) 
greater than 0.9 MeV are used for both detectors. 
For a comparison, we also show the sensitivity, at 90\% CL, expected 
for SADO$_{\rm multi}$ if the contributions of all 16 NPPs were 
included.}
\label{kl-sado-90-kashi}
\end{figure}

To answer this question, we present in Fig.~\ref{kl-sado-90-kashi} 
a comparison between the sensitivities to $\sin^2\theta_{12}$ 
attainable by SADO$_{\rm single}$, SADO$_{\rm multi}$ located at 54 km, 
and KamLAND. 
This plot shows how accurately we can reproduce the input 
mixing parameters, $\sin^2 \theta_{12}= 0.29$ and 
$\Delta m^2_{21} = 7.9 \times 10^{-5}$ eV$^2$, which will be
used throughout this paper.

For the purpose of comparison we have used for both, SADO and  
KamLAND, the same systematic error of 4\% and the common 
energy threshold of $E_{\text{prompt}} = 0.9$ MeV.
We determined the exposure time of the each detector so that 
they receive (approximately) the same number of events without 
oscillation.
By equalizing the number of events in this comparison, 
we aim to reveal how efficient the principle of tuning 
to the oscillation maximum for determining $\theta_{12}$ is.
For SADO$_{\rm multi}$ and KamLAND we have included all 16 NPPs, 
while for SADO$_{\rm  single}$ we have considered only contributions 
from the Kashiwazaki-Kariwa NPP. 
(The distances to the 16 NPPs from SADO and KamLAND are given in Table~\ref{Tabzero}.)

In comparing KamLAND with SADO in this subsection, we have ignored 
the geo-neutrino contributions.
Therefore, the comparison between SADO$_{\rm multi}$ and KamLAND 
is purely a comparison of the two different detector locations. 
The possible impact of geo-neutrinos on SADO will be described in 
Sec.~\ref{requirement}, and details of the analyzes in 
Secs.~\ref{procedure} and \ref{analysis}.

We discover in Fig.~\ref{kl-sado-90-kashi} the enormous power of SADO;
the sensitivity attainable by SADO is better by a factor of $\simeq$
2.5 than that of KamLAND for about {\it three times} longer operation 
in kt$\cdot$yr.
We also notice that the sensitivities achievable by SADO$_{\rm
single}$ and SADO$_{\rm multi}$ are comparable; the additional
contribution from the other 15 reactors does not affect so much the
$\theta_{12}$ determination.  %
More quantitatively, we recognize from
Fig.~\ref{kl-sado-90-kashi} that at 90\% CL, the sensitivity
attainable by SADO$_{\rm single, multi}$ corresponds to about 7\%
(5\%) error in $\sin^2\theta_{12}$ for 0.5 kt$\cdot$yr (1.5
kt$\cdot$yr) measurement.

One may want to ask a further question. Namely, how long 
does KamLAND take to reach the same sensitivity as of SADO?
(In fact, this is the question raised by one of the referees.) 
The answer to the question is in fact a surprising one: 
To achieve the same sensitivity of SADO$_{\rm multi}$ presented 
in the upper panel of Fig.~\ref{kl-sado-90-kashi}, KamLAND would 
need more than 100 kt$\cdot$yr exposure.
Whereas for the cases SADO$_{\rm single}$ shown in the upper panel  
and SADO$_{\rm single,multi}$ shown in the lower panel, KamLAND 
would not achieve the same sensitivities even in the limit of 
infinite statistics. Therefore, tuning the baseline distance is essential 
to achieve the ultimate sensitivity of 2\% level for 
$\sin^2\theta_{12}$ determination.

We should mention that, for the lower value of $\Delta
m^2_{21}=7.2\times 10^{-5}$ eV$^2$, preferred by the
data~\cite{KamLAND} before KamLAND 2004 result, the optimal distance
would have to be re-scaled to about 70 km (see Eq.(\ref{eq:scale})),
resulting in a corresponding increase of a factor of two in running
time in SADO to match the same sensitivity at higher $\Delta m^2_{21}$
achieved for 54 km.

\subsection{$\sin^2{\theta_{12}}$ versus $\sin^2{2\theta_{12}}$}

Though it is slightly off the main line of the discussion 
let us make some comments on usage of the variable in the 
plots we present in this paper. 
There have been various choices of variables for displaying 
allowed regions in $\theta_{12}$ - $\Delta m^2_{21}$ plane. 
They include,  $\tan^2{\theta_{12}}$, $\sin^2{2\theta_{12}}$, and 
$\sin^2{\theta_{12}}$. We feel it important, if possible, 
to have a ``standard choice'' of the mixing variable to make 
comparison between different analysis easier. 
We choose $\sin^2{\theta_{12}}$ throughout this paper for 
reasons which are described below.

First of all, we now know that $\theta_{12}$ is in the first octant. 
Therefore, we have no more reasons to continue to use 
$\tan^2{\theta_{12}}$. Then, the question is which of 
$\sin^2{2\theta_{12}}$, or $\sin^2{\theta_{12}}$ is to be used. 
This is a difficult choice because both of them can be 
supported for good reasons from physics point of view.
The solar neutrinos experience neutrino flavor transformation 
by different mechanisms at low and high energies. 
At low energies, it is essentially the vacuum oscillation 
effect that is dominant. At high energies the adiabatic MSW 
conversion effect takes over where the $\nu_e$ survival 
probability is approximately given by $\sin^2 \theta_{12}$. 
Therefore, the natural variable at low and high energies are 
$\sin^2{2\theta_{12}}$, and $\sin^2{\theta_{12}}$, respectively, 
and they both have right to be chosen.

We use $\sin^2{\theta_{12}}$ in this paper because of two reasons. 
One is that $\sin^2{\theta_{12}}$ allows a direct physical interpretation 
as the probability of finding $\nu_{e}$ in the second mass eigenstate, 
as emphasized in \cite{parke-mena}. 
The other one is that $\theta_{12}$ is relatively large, though not quite maximal. 
For large mixing angle $\theta$, $\sin^2{2\theta}$ is not 
a convenient variable because $\sin^2{2\theta}$ becomes 
insensitive to change in $\theta$. 
It can be simply understood by computing the Jacobian
\begin{eqnarray}
\frac{ \delta (\sin^2{2\theta}) }{\delta(\sin^2\theta)} \simeq
\frac{d \sin^2{2\theta}}{d \sin^2\theta} = 
4 \cos{2\theta},  
\label{jacobian}
\end{eqnarray} 
where $\delta (\sin^2\theta)$ and $\delta (\sin^2{2\theta})$
denote the error in $\sin^2\theta$ and $\sin^2{2\theta}$, respectively.
It means that a large error obtained for $\sin^2\theta$ can be translated 
into a much more suppressed error for $\sin^2{2\theta}$ 
because of the small Jacobian at large $\theta \sim \pi/4$.
Of course, the problem of disparity between errors estimated for 
these two variables in our case is much milder than that of 
atmospheric angle $\theta_{23}$ \cite{MSS}.

\section{Requirements and possible obstacles for the precision measurement of $\theta_{12}$}
\label{requirement}

In order to achieve the optimal sensitivity promised in the previous sections 
the following two requirements have to be met. They are:
(i) experimental systematic error of 4\% level, 
(ii) baseline distance of 50-70 km. 
We will discuss the requirement (i) and other related issues in this section. 
We calculate the optimal distance after setting up our analysis procedure 
in the next section.

\subsection{Experimental systematic error}
\label{subcec:systematic}

The key ingredients in the accurate measurement of $\theta_{12}$ 
in reactor experiments is the systematic error in the measurement. 
In this paper, we assume that there exists a dedicated front detector. 
It would be ideal if there already exists a near-far detector complex 
for reactor measurement of $\theta_{13}$ at less than 
1-2 km from the principal reactor for SADO. 
Presence of the front detector is crucial to suppress the systematic errors 
for accurate measurement of $\theta_{12}$ because the various errors such 
as absolute flux and cross sections largely cancel between the front detector 
and SADO, in complete parallelism with the reactor $\theta_{13}$ 
measurement.  
See e.g., \cite{munich,reactorCP} for details on 
how and which errors can be canceled.

Since SADO cannot have structure exactly identical with the 
front detector, the systematic error may not be as small as 
$\sim$ 1\%, the number expected to be reachable in reactor $\theta_{13}$ 
measurement \cite{reactor_white}. 
We assume in our analysis, based on \cite{suekane}, that the systematic error 
can be as small as 4\% at SADO.\footnote{
In the first and the second versions of this paper posted on the 
arXiv (hep-ph/0407326), we have used 
$\Delta m^2_{21} = 8.2 \times 10^{-5}$ eV$^2$ based on the earlier version 
of \cite{KamLAND_new}, which is 3.8\% larger than the value used in 
the third and this published (fourth) version. 
The readers who want to examine how different are the results with 
slightly different values of $\Delta m^2_{21}$ can look into 
the second version of the paper. 
More importantly, we assumed systematic error of 2\% (4\%) in the first 
(second, third and fourth) version. 
Therefore, the readers who want to do a detailed comparison 
between the cases of systematic error of 4\% and 2\% are advised to look into 
the first version of the paper. 
The reason why the sensitivities to $\theta_{12}$ are essentially 
unchanged is that the measurement is not systematics dominated even at 
60 GW$_{\text{th}}\cdot$kt$\cdot$yr.}
This systematic error is supposed to include background from
cosmic-ray events, assuming that the overburden is enough.
Our claim that 4\% systematic error may be in reach can be regarded as 
reasonable 
by recalling that  the CHOOZ experiment \cite{CHOOZ} already achieved 
the goal of the systematic error (1.5\% for detection efficiency error) 
apart from the one due to flux normalization.
Therefore, we think that the 4\% is a conservative estimate and a smaller 
systematic error of $\simeq$2\% may well be thinkable.

One of the consequence for the requirement of less than 4\% systematic error 
is that we cannot rely on the event cut of the energy spectrum of prompt energy 
$E_{\text{prompt}} > 2.6$ MeV  to reject geophysical neutrino contamination, 
as done by the KamLAND group in their reactor neutrino analysis 
\cite{KamLAND,KamLAND_new}.
Because of the uncertainty in the energy measurement it produces an additional 
$\simeq 2\%$ systematic error which makes over-all 4\% systematic error 
difficult to achieve. Therefore, we must take all the events with prompt energy 
greater than 0.9 MeV. 
But it means that we have to deal with geophysical neutrinos in our analysis. 
It will be one of the most important issues in our sensitivity analysis in 
this paper.

\subsection{Geophysical neutrinos; a brief review}
\label{geo-sec}

Geo-neutrinos comes into play in our sensitivity analysis of $\theta_{12}$ 
because U (Th) decay series processes through six (four) $\beta$-decays, 
producing 6 (4) geo-neutrinos with energy $E_\nu \lsim 3.4$ MeV, or 
$E_{\text{prompt}} \lsim 2.4$ MeV. 
The threshold of $E_\nu^{\text{th}}=1.7$ MeV at SADO then 
implies that geo-neutrino events from U and Th decays can contaminate 
reactor neutrino events in our analysis. 

Then, what are geo-neutrinos and do we know their properties?
In the rest of this subsection, we briefly review what we know about geophysical 
neutrinos.
The radioactive decay chains of radionuclides such as 
$^{40}$K, $^{238}$U and $^{232}$Th inside the Earth produce not only heat 
but also electron antineutrinos. The fact that the radioactive heat flux and 
the geophysical neutrino flux are
tightly linked was first pointed out by Eder~\cite{eder} in the sixties. 
Since then, the geo-neutrinos have been the subject of 
continuing interests~\cite{geonu}. 
Unfortunately, a quantitative accounting of the radioactive heat flux
requires a detailed knowledge of the abundance distributions of
these long-lived radioactive species inside the Earth.
Our current knowledge of these abundance distributions is incomplete 
because of many reasons, for example, direct sampling is 
possible only at or near the surface and most of the Earth's surface is 
inaccessible. Consequently, the radiogenic heat flux and therefore the 
geo-neutrino flux is largely model-dependent and its precise magnitude 
is unknown.

The presently estimated value for the global heat flux from the earth is 40 
TW. It gives an upper bound on geo-neutrino flux from the interior of the 
earth. Heating from known radioactive sources in the surface layers 
(corresponding to 1/300 of the Earth's total mass) is estimated to be
already about 20 TW. 
Most geophysical models predict that the concentrations of radioactive 
sources will decrease rapidly with depth, but even small variations in 
these concentrations, for such a big mass as that of our planet, can greatly 
affect heat and geo-neutrino production.

We will treat geo-neutrinos flux as an additional parameter to be fit 
in our sensitivity analysis. 
To take proper care of its uncertainty we examine the two extreme cases of 
no geo-neutrino flux as an input and the maximum flux corresponding to 
global heat flux of 40 TW (the Fully Radiogenic Model). 
Fortunately, as we will see in the next two sections, 
the sensitivity on $\theta_{12}$ to be achieved by SADO is not 
disturbed in an essential way by the presence of geo-neutrinos
provided that we choose appropriate baseline distances.

\subsection{Uncertainty due to $\theta_{13}$}

Uncertainty due to the unknown value of $\theta_{13}$ can be another 
source of the systematic error. It was shown in \cite{concha-pena} that 
it can be regarded as an effective uncertainty in the flux normalization. 
Let us briefly review their argument. 
Since the vacuum neutrino oscillation is a good approximation 
for the reactor $\theta_{12}$ experiments, 
we can write down the electron antineutrino survival 
probability in a simple approximate form 
\begin{eqnarray}
P(\bar{\nu}_e\leftrightarrow\bar{\nu}_e) =   
\sin^4\theta_{13} + \cos^4\theta_{13} 
\left[ 1 - \sum_i f_i \sin^22\theta_{12} 
\sin^2\left(\frac{\Delta m^2_{21}L_i}{4 E_{\nu}}\right)\right],
\label{prob}
\end{eqnarray} 
where $E_{\nu}$ is the neutrino energy, 
$L_i$ is the distance to the detector from reactor $i$, and 
$f_i$ is the fraction of the 
total neutrino flux which comes from reactor $i$. 
Thus, nonzero $\theta_{13}$ effectively acts as an uncertainty in 
the flux normalization of less than 8\%, giving the CHOOZ constraint 
$\sin^22\theta_{13} < 0.15$ at 90\% CL \cite{bari_update}.

This uncertainty in the flux normalization cannot be canceled by 
measurement at the front detector. Therefore, $\theta_{13}$ has 
to be determined by an independent measurement. 
But the experimental determination of $\theta_{13}$ should always 
come with errors, $\delta (\sin^2{2\theta_{13}})$. 
It will produce an effective uncertainty in the probability by the amount 
\begin{eqnarray}
\delta (\cos^4\theta_{13}) 
\simeq 
\left|
\frac{d \cos^4\theta_{13}} {d (\sin^2{2\theta_{13}})} 
\right|
\delta (\sin^2{2\theta_{13}}) 
\simeq
\frac{1+\tan^2{\theta_{13}}}{2} \delta (\sin^2{2\theta_{13}}),
\label{jacobian2}
\end{eqnarray} 
for a given experimental error $\delta (\sin^2{2\theta_{13}})$ of 
measurement.
If we take the estimation in \cite{MSYIS} for reactor $\theta_{13}$ 
experiment, $\delta (\sin^2{2\theta_{13}}) \simeq 0.012$ 
(0.02 at 90\% CL) almost 
independently of the true values of $\sin^2{2\theta_{13}}$.
Therefore, we expect an uncertainty of $\simeq 0.6$\% level in the 
probability and hence in $\sin^2{2\theta_{12}}$. 
It may be translated, by using (\ref{jacobian}), into an uncertainty 
on $\sin^2{\theta_{12}}$ of about 0.4\% at the current best fit 
of $\theta_{12}$.

Thus, the error of $\theta_{13}$, once it is measured by reactor
experiments, adds only $\simeq$ 0.4\% to the uncertainty on
$\sin^2{\theta_{12}}$.  Though it does not pose any serious problem,
more precise measurement of $\theta_{13}$ either by LBL high-intensity
superbeam experiments with both neutrino and antineutrino modes
\cite{JPARC,NuMI,SPL}, or by high statistics reactor
experiments~\cite{reactor_white} or even by neutrino factory
\cite{nufact} is highly welcome.

We assume in the rest of this paper vanishing $\theta_{13}$ for simplicity. 
Assuming $\theta_{13}$ is determined with error much smaller than 1\%  
the effect of nonzero $\theta_{13}$ effectively acts as smaller 
$\bar{\nu}_{e}$ flux by a factor of $\cos^4{\theta_{13}}$. 
For example, if $\theta_{13}$ is on the edge of the above CHOOZ bound,  
$\cos^4{\theta_{13}} = 0.92$, we need 9\% longer 
running time to obtain the same sensitivities as presented in 
the later sections.

\section{Analysis procedure and estimation of the optimal distances}
\label{procedure}

In this section, we first make a rough description of our analysis procedure, 
and then estimate the optimal distances for reactor $\theta_{12}$ experiments. 
For more details of the analysis procedure, we refer the readers to Appendix~\ref{ap2}.

\subsection{Analysis procedure in brief}

We follow our previous papers~\cite{NTZ02,Nunokawa:2003dd} in calculating 
the number of neutrino events expected from reactors as well as 
from Th and U radioactive decays in the Earth at a liquid scintillator detector, SADO.
We will give  our results in terms of 
$\text{GW}_{\text{th}}\cdot \text{kt} \cdot\text{yr}$.
Notice that 
10 $\text{GW}_{\text{th}}\cdot \text{kt} \cdot\text{yr }$
is approximately equivalent to 1 year exposure of 
0.5 $\text{kt}$ detector at 54 km away from Kashiwazaki-Kariwa NPP 
operating at 80\% efficiency. 

The total number of events from a single NPP for an exposure of 
1 $\text{GW}_{\text{th}}\cdot \text{kt} \cdot\text{yr }$,  
for the used threshold energy is estimated to be, 

\begin{equation}
N^{\text{reac}}_{\text{total}} (E_{\text{prompt}} > 0.9 {\text{ MeV}}) = 
\displaystyle \left \{
\begin{array}{c}
 141 \nonumber\\  46  
\end{array} 
\right \} \times
\left[\frac{L}{54\ \text{km}}\right]^{-2} \ \ 
[\text{GW}_{\text{th}}\cdot \text{kt} \cdot\text{yr }]^{-1},
\end{equation} 
where the upper (lower) value corresponds to no oscillation (oscillation at 
the input values used in this work).    
$\GWth$ is defined to be the thermal power actually generated, which should
not be confused with the maximal thermal power of a given NPP. 

The total number of events from geo-neutrinos expected from the 
Fully Radiogenic Model~\cite{geonu} at Mt. Komagatake is 
\begin{equation}
N^{\text{geo}}_{\text{total}} (E_{\text{prompt}} > 0.9 {\text{ MeV}}) = 49 
\ [\text{kt} \cdot\text{yr }]^{-1}.
\end{equation} 

Since we currently do not know the U and Th geo-neutrino fluxes we 
consider two extreme cases: 
(i) zero geo-neutrino input and 
(ii) geo-neutrino input flux from U and Th computed assuming that the 
radiogenic production accounts for the total Earth heat flow of 40 TW 
(input from the Fully Radiogenic Model). 
They represent two extremes of the input geo-neutrino flux between 
zero and 3$\times 10^{7}$ cm$^{-2}$ s$^{-1}$ at the position of 
the detector, and one would expect the real situation to be somewhere 
between these two numbers. 
Note that even if we assume zero geo-neutrino input, we allow non-zero 
geo-neutrino flux as an output in our fit which can confuse reactor neutrinos. 
Since the ratio of U to Th contributions is 
rather well defined and common in varying models, we assume in our analysis 
that their relative contribution is held fixed while we treat an over-all 
absolute normalization as a free parameter. The Th contribution is assumed 
to be 83\% of the total geo-neutrino flux~\cite{geonu}. 
This procedure should be (and will be) tested by future KamLAND data 
which includes $E_{\text{prompt}} < 2.6$ MeV.
Of course, having the additional free parameter in general makes the 
sensitivity on  $\theta_{12}$ worse. 

We use the following definition of the $\chi^2$ function as
  \begin{equation}
  \chi^2 (\sin^2 \theta_{12},\Delta m^2_{21}, \phi_{\text{U}})
= \displaystyle \sum_{i=1}^{17} 
  \frac{\left[ 
N_i^{\text{exp}}-
N_i^{\text{theo}}(\sin^2 \theta_{12},\Delta m^2_{21},\phi_{\text{U}})\right]^2}
{\sigma_i^2},
\label{def_chi2}
  \end{equation}
where $\sigma_i = \sqrt{ N_i^{\text{exp}}+(0.04\, N_i^{\text{exp}})^2}$ 
is the statistical plus systematic uncertainty in the number of 
events in the $i$-th bin; 
$N_i^{\text{theo}}$ is the theoretical expected number of events
as functions of mixing parameters as well as $\phi_{\text{U}}$, 
the total geo-neutrino flux from U, to be fitted, 
calculated as explained in the Appendix~\ref{ap2};  
$N_i^{\text{exp}} \equiv 
N_i^{\text{theo}}(\sin^2 \theta_{12}(\text{input}),
\Delta m^2_{21}(\text{input}), \phi_{\text{U}}(\text{input}))$ 
is the simulated expected number 
of events at the detector for the input values of the mixing parameters 
as well as geo-neutrino flux, assumed in this paper, {\em i.e.}, 
$\sin^2 \theta_{12}({\text{input}})= 0.29$ 
and  $\Delta m^2_{21}({\text{input}}) = 7.9 \times 10^{-5}$ eV$^2$ and 
the geo-neutrino input, $\phi_{\text{U}}(\text{input})$, 
will be either zero or equal to what is expected if   
the radiogenic contribution to the terrestrial heat is to be 
explained by the Fully Radiogenic Model. 
The crustal contribution was estimated as in  Ref.~\cite{Nunokawa:2003dd} but for 
a detector located at Mt. Komagatake in Niigata prefecture, Japan.\footnote{
Be careful about the fact that there are at least 13 Mt. Komagatake in Japan.}

Using the $\chi^2$ function we compute the region in the 
$(\sin^2 \theta_{12}, \Delta m^2_{21})$ space allowed by 
SADO spectrum data at 68.27\%,
90\%, 95\%, 99\%  and 99.73\% CL for a given
$\GWth \cdot$kt$\cdot$yr exposure.

\subsection{Optimal baseline distance}

We have discussed in Sec.~II that it is the key for a precise measurement 
of $\theta_{12}$ measurement to choose the baseline 
distance which corresponds to the oscillation maximum. 
To seek ultimate sensitivity on  $\theta_{12}$, however, we must first 
elaborate this point.

\begin{figure}[htbp]
\vglue -.1cm
\begin{center}
\includegraphics[width=0.7\textwidth]{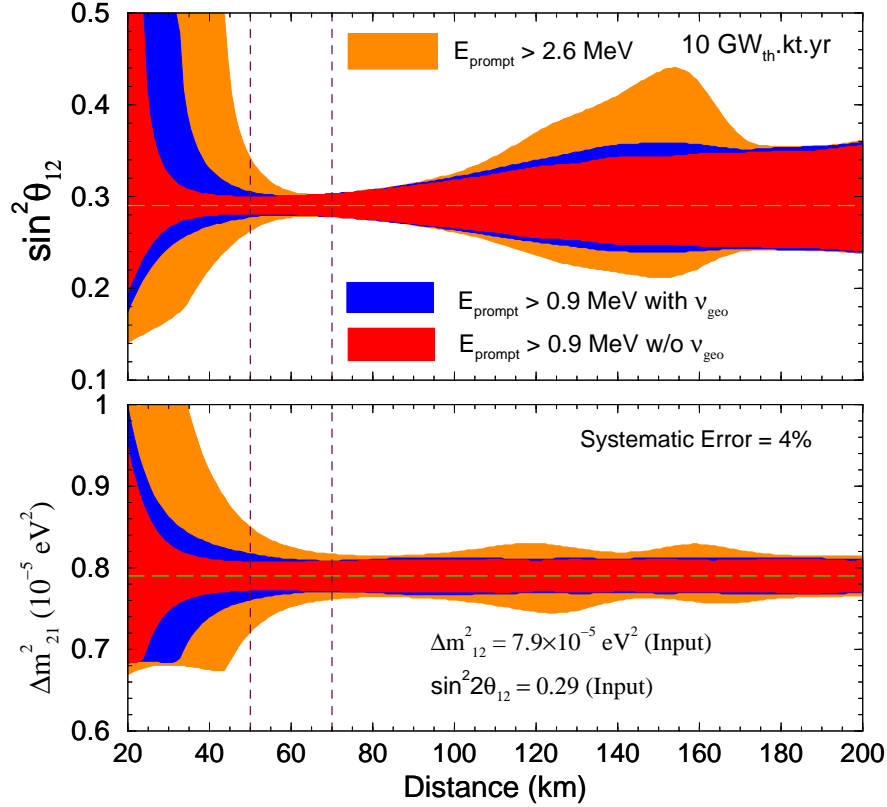}
\end{center}
\vglue -0.3cm
\caption{
Achievable precision of the determination of the 
oscillation parameters at 68.27\% CL for 1 DOF 
as a function of the detector distance 
from the Kashiwazaki-Kariwa NPP for 
10 GW$_{\text{th}}\cdot$kt$\cdot$yr exposure and  energy 
thresholds of $0.9$ and 2.6 MeV. 
The horizontal dashed lines indicate the input values of 
the mixing parameters.
The geo-neutrino contribution was calculated as 
 predicted by the Fully Radiogenic Model.
For the lower energy threshold the $\chi^2$ was minimized 
allowing the total geo-neutrino flux to vary from 0 to 3$\times 10^{7}$ cm$^{-2}$ s$^{-1}$. 
We have used the input values of the mixing parameters 
corresponding to  $\sin^2 \theta_{12}= 0.29$ and 
$\Delta m^2_{21} = 7.9 \times 10^{-5}$ eV$^2$.
The vertical dashed lines indicate the optimal range for 
SADO detector. 
}
\label{dist-dep}
\end{figure}

We show, in the upper and the lower panels in Fig.~\ref{dist-dep}, 
the accuracy of determination of $\sin^2\theta_{12}$ and $\Delta m^2_{21}$,  
respectively, as a function of $L$ expected by 
10 GW$_{\text{th}}\cdot$kt$\cdot$yr 
exposure  computed by the minimization of 
the $\chi^2$ function that is defined in Sec. III.
The results for two different energy thresholds, $0.9$ and $2.6$ MeV, 
are presented for the purpose of comparison.  
For the case $E_{\text{prompt}}>0.9$ MeV, we show two curves 
for sensitivity with and without geo-neutrino background contribution. 
We note that both $\Delta m^2_{21}$ and $\theta_{12}$ are left free 
and marginalized in the analysis to obtain the sensitivity of 
$\sin^2\theta_{12}$ and $\Delta m^2_{21}$, as presented 
respectively in the upper and the lower panels in Fig.~\ref{dist-dep}.

We see that when the analysis threshold is higher the best place to put  
a detector, in order to achieve greater sensitivity to $\theta_{12}$,  
is at around $L\simeq L_{\text{OM}}  = 63$ km,  
in agreement with our rough expectation. 
On the other hand, when the threshold is lower, it is preferable to 
choose values of $L$ 10-25\% smaller than $L_{\text{OM}}$\footnote{
This fact was recognized in a careful study of sensitivity in reactor 
$\theta_{13}$ measurement \cite{reactor_white,munich}. 
But, the shorter baseline $L$ by 10-25\% does not make a great difference 
in that case because $L$ is short, $\sim 1$ km. 
But  in reactor $\theta_{12}$ experiments, however, the difference is 
significant because the choice of $L=50$ km or $L=70$ km 
implies a completely different site for  SADO.}.

It is remarkable that if there were no geo-neutrinos the baseline 
distance as short as 40 km could be suitable. 
Even when we take them into account, 
the optimal distance ranges between 50 and 70 km 
which is still a relatively wide range. 
We will see below (Sec.~\ref{analysis}) that even at 70 km, 
the longest end of the distance range, the sensitivity is worse 
only by about 15\% or so than SADO at 54 km. 
This point is very important and encouraging for people who try 
to design reactor $\theta_{12}$ experiments because a wide band of 
radii between 50 and 70 km around a particular reactor NPP
can serve as a potential location for the experiment.

The reason why the effect of geo-neutrinos is larger at shorter distances 
is as follows. 
As $L$ gets shorter the region where oscillation effect is sizable moves to 
lower energies. Then,  for $L \lsim 50$ km it starts to invade the region of 
$E_{\nu} < 3.4$ MeV  where geo-neutrinos live.

An interesting behavior of the error on  $\sin^2\theta_{12}$
for the $2.6$ MeV threshold, rapidly increasing with $L$ 
from about 80 to 150 km and decreasing again from 150 to about 170 km,  
can be explained as follows.  The maximal sensitivity, reached 
at about 60 km, corresponds to the first oscillation maximum at the 
peak of the event distribution. As $L$ increases this first oscillation
maximum moves to higher energies, finally going out of the reactor 
neutrino energy spectrum. When the second oscillation maximum passes 
the  $2.6$ MeV threshold, at about 150 km, it improves the  
$\sin^2\theta_{12}$ determination causing the partial recovery of 
precision.

In the lower panel of Fig.~\ref{dist-dep} we appreciate the 
weak dependence on $L$ of the accuracy in 
the determination of $\Delta m^2_{21}$ for $L\gsim$ 50 km.
We will see in Sec.~\ref{comparison} (Table~\ref{Tabmass}) that 
SADO does improve the sensitivity of $\Delta m^2_{21}$ determination 
over the KamLAND's though not to the extent that occurs for $\theta_{12}$.   

It is interesting to clarify the difference between our case and that 
of the experimental set up of the next generation LBL
accelerator neutrino experiments \cite{JPARC,NuMI,SPL}. 
In particular, in the JPARC-SK experiment, the accuracy of
determination of $\sin^2{2\theta_{23}}$ is expected to reach
down to $\simeq$ 1\% at 90\% CL \cite{JPARC}.
In this experiment, where $L$ is already fixed, the best sensitivity 
for the angle determination is obtained when the experiment 
is tuned to the energy which corresponds to the 
first oscillation maximum. The spectrum of an off-axis beam is nearly monochromatic.
On the other hand, in the reactor experiment we are considering, 
the energy spectrum is given and is quite broad and hence 
nearly maximal sensitivity to the mixing angle can be achieved in a range of  
values of $L$.

\section{How accurate is the reactor measurement of $\theta_{12}$?} 
\label{analysis}

In this section, we give a quantitative estimate of the sensitivity 
which can be achievable in a dedicated reactor measurement of $\theta_{12}$.

\subsection{Analysis of sensitivity on $\theta_{12}$: SADO at 54 km}

\begin{figure}[htbp]
\begin{center}
\vglue -0.6cm
\includegraphics[width=1.2\textwidth]{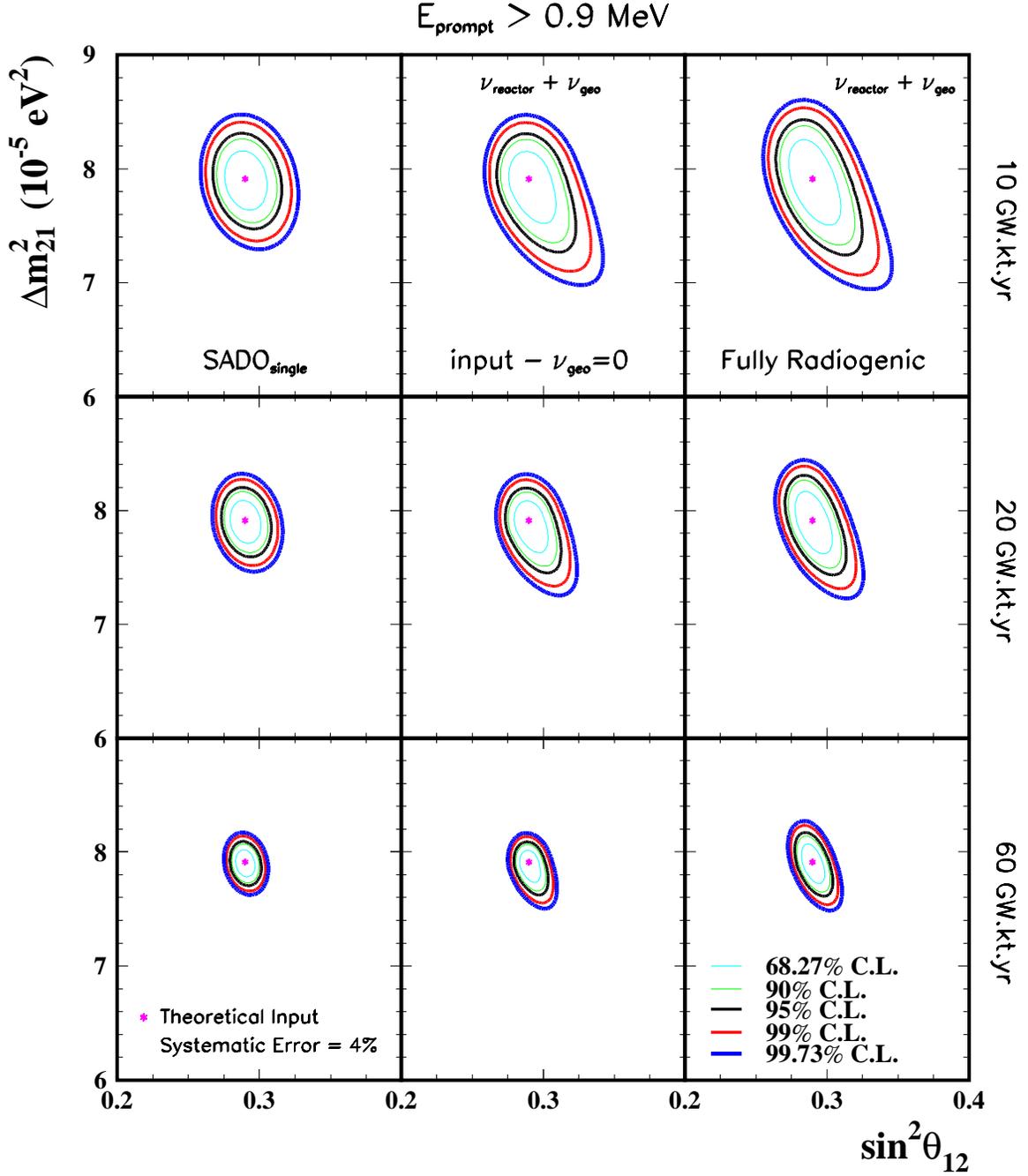}
\end{center}
\vglue -1.7cm
\caption{
Precision of the determination of the 
oscillation parameters by SADO$_{\text{single}}$, a detector at 54 km 
from the Kashiwazaki-Kariwa NPP, for exposures of 
10 $\GWth \cdot$kt$\cdot$yr (upper panels), 
20 $\GWth \cdot$kt$\cdot$yr  (middle panels) 
and 60 $\GWth \cdot$kt$\cdot$yr  (lower panels). Left and  middle panels 
have no geo-neutrino input, but in the middle panels geo-neutrinos were 
however taken into account in the fit. Right panels were calculated for 
geo-neutrino input according to the Fully Radiogenic Model.  
The confidence level regions were computed for 2 DOF. 
}
\label{sado54a}
\end{figure}

\begin{figure}[htbp]
\begin{center}
\vglue -0.6cm
\includegraphics[width=1.2\textwidth]{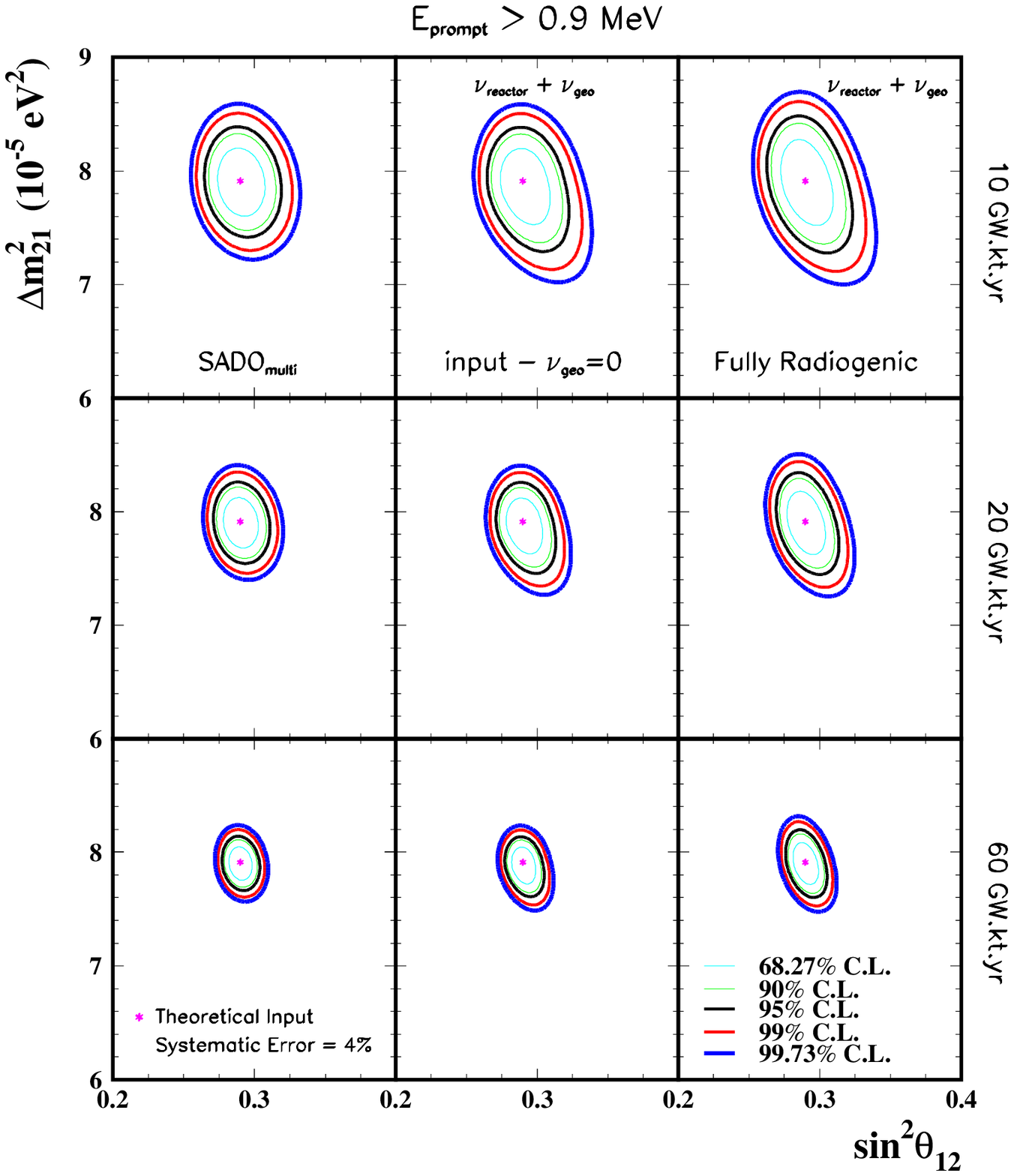}
\end{center}
\vglue -1.7cm
\caption{
Same as Fig.~\ref{sado54a} but for SADO$_{\text{multi}}$ with all 16 reactor NPPs included. 
}
\label{sado54b}
\end{figure}

We show in Fig.~\ref{sado54a} how precisely the mixing parameters 
can be determined by SADO$_{\text{single}}$, assuming that only the 
Kashiwazaki-Kariwa NPP will be contributing to the neutrino flux at the 
detector site, for exposures of 
10 $\GWth \cdot$kt$\cdot$yr (upper panels), 
20 $\GWth \cdot$kt$\cdot$yr  (middle panels) 
and 60 $\GWth \cdot$kt$\cdot$yr  (lower panels).
In the left panels, geo-neutrinos were not considered, in the 
middle panels, we have set the geo-neutrino flux input to be zero but allowed 
the flux to be fitted to vary freely in the $\chi^2$ fit. 
In the right panels, we have calculated the geo-neutrino 
input according to the Fully Radiogenic Model and allowed 
the flux to be fitted to vary freely in the $\chi^2$ fit.   

In Fig.~\ref{sado54b} we provide the same information as in 
Fig.~\ref{sado54a} but for SADO$_{\text{multi}}$ which includes  
the contributions of all the 16 reactor NPPs in Japan. 
We note that by taking into account the contributions from 
the other 15 reactor NPPs, the accuracy of determination of 
$\sin^2\theta_{12}$ becomes slightly worse. The error on 
$\Delta m^2_{21}$ also slightly increases.

\begin{figure}[htbp]
\vglue -3.0cm
\begin{center}
\includegraphics[width=0.8\textwidth]{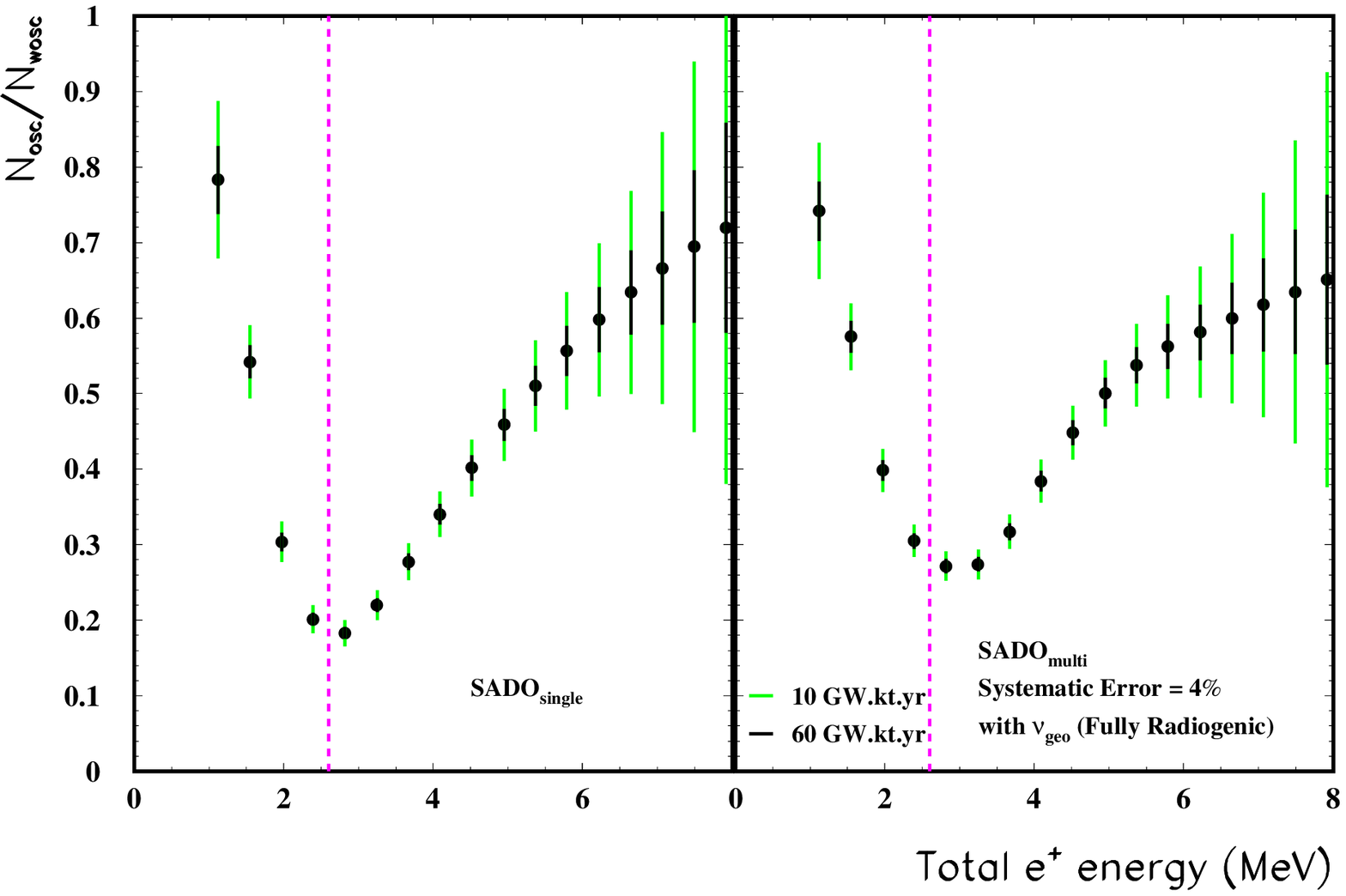}
\end{center}
\vglue -2.0cm
\caption{Spectral distortion expected as a function of the $e^+$ 
prompt energy for SADO$_{\text{single}}$ (left panel) and 
SADO$_{\text{multi}}$ (right panel).
Each ratio is represented as a point with the corresponding error
bar for a given exposure (10 and 60 $\GWth \cdot$kt$\cdot$yr).
The number of oscillation events at each bin was 
calculated for the input values $\sin^2 \theta_{12}= 0.29$ and 
$\Delta m^2_{21} = 7.9 \times 10^{-5}$ eV$^2$.
The dashed line indicates the maximal energy for geo-neutrinos. 
Geo-neutrino contribution from the Fully Radiogenic Model was 
included. 
}
\label{spec}
\end{figure}

The inclusion of geo-neutrinos does not influence very much 
the sensitivity to $\sin^2\theta_{12}$
as can be seen in Figs.~\ref{sado54a} and \ref{sado54b}. 
This is because we have chosen a distance such that the 
position of the dip is far enough from the geo-neutrino energy 
region. This can be visualized in Fig. \ref{spec}, where we present 
the  normalized expected energy spectra at SADO$_{\text{single}}$ 
(left panel) and SADO$_{\text{multi}}$ (right panel) for exposures 
of 10 and 60 $\GWth \cdot$kt$\cdot$yr.
The impact of the inclusion of the 15 NPPs is to modify somewhat 
the shape of the spectrum.

\begin{figure}[htbp]
\begin{center}
\includegraphics[width=0.65\textwidth]{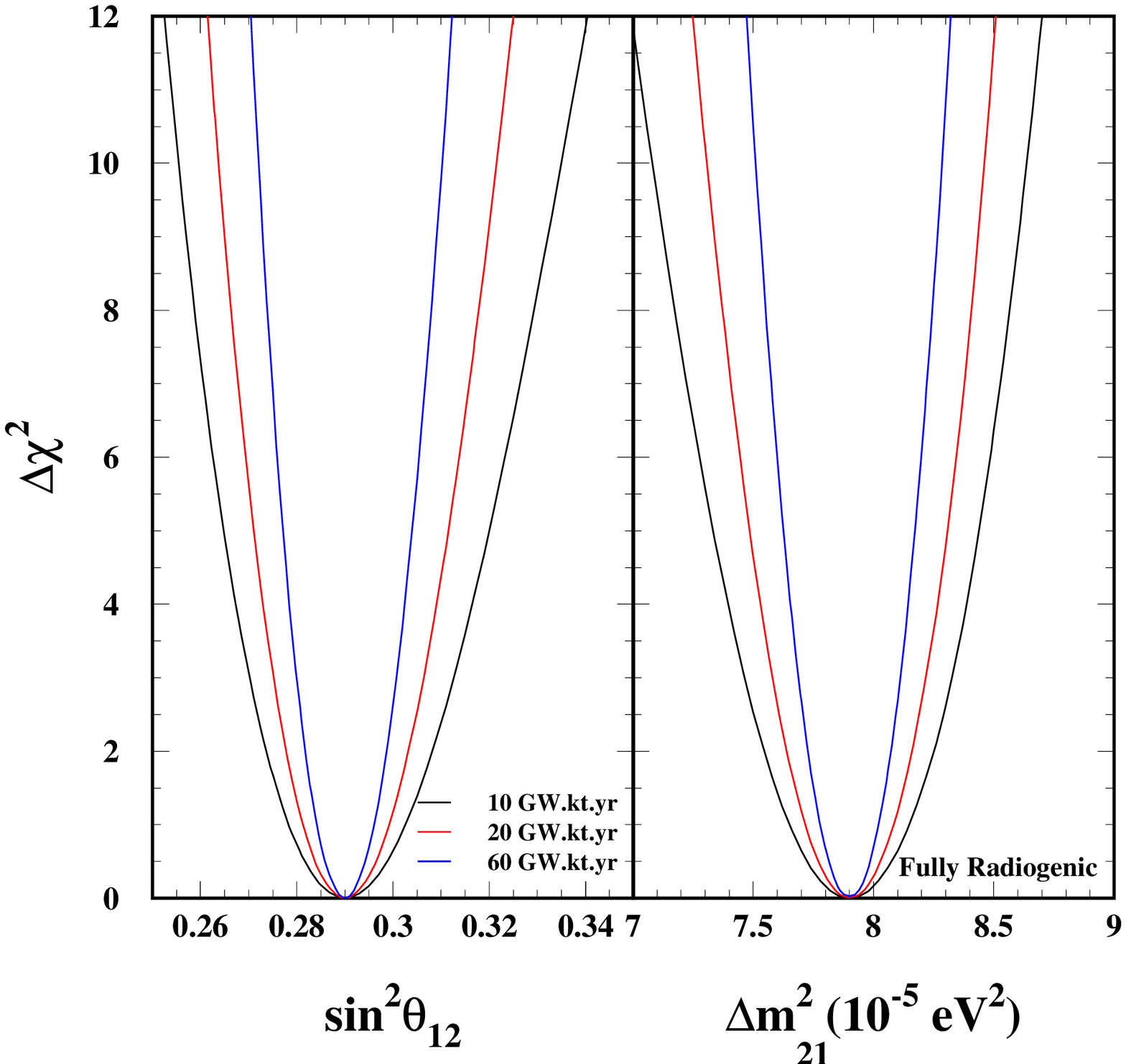}
\end{center}
\vglue -0.5cm
\caption{
$\Delta \chi^2$ as a function of the oscillation parameters around 
the minimum for 10, 20 and 60 $\GWth \cdot$kt$\cdot$yr exposure. 
All 16 reactor NPPs were included in the calculation and 
the prompt energy threshold used was 0.9 MeV. Geo-neutrino 
contribution from the Fully Radiogenic Model was included.
}
\label{chi2}
\end{figure}

On the other hand, the sensitivity to $\Delta m^2_{21}$ is slightly more 
affected by the inclusion of the geo-neutrino background. 
This is because in  $\Delta m^2_{21}$ determination, it is important to 
observe the shape of the spectrum in the entire energy range making it more 
sensitive to the geo-neutrino contributions in the lower energy bins.
At this point it is also instructive to look at Fig.~\ref{chi2}, 
where we show the $\Delta\chi^2=\chi^2$ behavior as a function 
of $\sin^2 \theta_{12}$ and $\Delta m^2_{21}$ for 
10, 20 and 60 $\GWth \cdot$kt$\cdot$yr exposure, all 16 NPPs considered. 

\begin{table}
\caption[aaa]{
The expected sensitivity to the solar mixing angle at
SADO$_{{\text{single}}}$ (SADO$_{{\text{multi}}}$)
at 1 $\sigma$ (68.27\%) and 3 $\sigma$ (99.73\%) CL for 1 DOF 
obtained for SADO at Mt. Komagatake (54 km, the upper rows) and 
Sado gold mine (71 km, the lower rows). 
They are calculated with the prompt energy threshold of 0.9 MeV and 
with the background effect of  geo-neutrinos from
Fully Radiogenic Model. 
The numbers for SADO$_{{\text{multi}}}$ are computed for both detectors 
by using position-dependent contributions from all the 16 NPPs.
}
\vglue 0.3cm
\begin{tabular}{c|cc}
\hline
\ exposure\  &\ \ $\sin^2 \theta_{12}$ [68.27\% CL] \
             &\ \ $\sin^2 \theta_{12}$ [99.73\% CL] \ \\
\hline
\multicolumn{3}{c}{54 km}\\
\hline
 10 $\GWth \cdot$kt$\cdot$yr
                 &\ \ $0.29^{+0.014}_{-0.013}$\ ($0.29^{+0.015}_{-0.014}$)
                 &\ \ $0.29^{+0.040}_{-0.031}$\
($0.29^{+0.042}_{-0.033}$)\\
\hline
 20 $\GWth \cdot$kt$\cdot$yr
                 &\ \ $0.29^{+0.010}_{-0.010}$\ ($0.29^{+0.011}_{-0.011}$)
                 &\ \ $0.29^{+0.028}_{-0.023}$\
($0.29^{+0.030}_{-0.025}$)\\
\hline
 60 $\GWth \cdot$kt$\cdot$yr
                 &\ \ $0.29^{+0.006}_{-0.006}$\ ($0.29^{+0.007}_{-0.007}$)
                 &\ \ $0.29^{+0.017}_{-0.015}$\
($0.29^{+0.019}_{-0.017}$)\\
\hline
\multicolumn{3}{c}{71 km}\\
\hline
 10 $\GWth \cdot$kt$\cdot$yr
                 &\ \ $0.29^{+0.016}_{-0.015}$\ ($0.29^{+0.019}_{-0.017}$)
                 &\ \ $0.29^{+0.045}_{-0.037}$\
($0.29^{+0.052}_{-0.040}$)\\
\hline
 20 $\GWth \cdot$kt$\cdot$yr
                 &\ \ $0.29^{+0.012}_{-0.011}$\ ($0.29^{+0.013}_{-0.012}$)
                 &\ \ $0.29^{+0.031}_{-0.027}$\
($0.29^{+0.036}_{-0.030}$)\\
\hline
 60 $\GWth \cdot$kt$\cdot$yr
                 &\ \ $0.29^{+0.007}_{-0.007}$\ ($0.29^{+0.008}_{-0.008}$)
                 &\ \ $0.29^{+0.018}_{-0.016}$\
($0.29^{+0.022}_{-0.020}$)\\
\hline
\end{tabular}
\label{Tab2}
\vglue 0.5cm
\end{table}

We summarize in Table~\ref{Tab2} the accuracy of determination of the 
mixing angle $\theta_{12}$ which can be achieved by SADO$_{\text{single}}$ 
and SADO$_{\text{multi}}$ for these three exposures.  
We use the prompt energy threshold of 0.9 MeV and take into account 
the background effect of geo-neutrinos expected by the Fully Radiogenic Model, the 
most conservative estimate.
We have investigated the dependence of this accuracy on the input 
value of $\Delta m^2_{21}$ by varying it in the range 
$\Delta m^2_{21} = (7-9) \times 10^{-5} {\text{ eV}^2}$. We have 
discovered the accuracy becomes somewhat worse, at most by 30\%,  
if the input value is smaller than the current best fit one 
due to geo-neutrinos, whereas  for higher values there is 
practically no change.

In Table~\ref{Tab2}, we also present for comparison the results of 
our analysis for a detector placed at Sado island, about 71 km 
away from the Kashiwazaki-Kariwa NPP.
As one can anticipate from Fig.~\ref{dist-dep} the sensitivity at 71 km is 
worse than that of SADO at 54 km but only by 15-20\%.
Thus, SADO at the Sado island is still a valid option for the reactor 
$\theta_{12}$ experiment.

\section{Stability of the results against changes in statistical procedure}
\label{stability}

We have shown in the previous sections that the sensitivity for 
$\theta_{12}$ attainable by SADO with a modest requirement of 
$\simeq$ 4\% systematic error is extremely good, which is competitive to 
all the methods so far proposed. 
In this section we confirm the stability of the results by using 
different definitions of $\chi^2$. 
While we believe that our statistical method and 
the treatment of errors based on $\chi^2$ defined 
in Eq. (\ref{def_chi2}) is reasonable and a standard one, 
it is nice if we can explicitly verify that our results 
are stable against changes in the statistical procedure.

For this purpose, we examine the following two possible choices of $\chi^2$.
The first one, which is frequently used in various statistical analyzes of data, 
reads 
\begin{equation}
\chi^2_{\text{corr}} = 
 \min_{\text{$\alpha$}}
\sum_{i=1}^{17}
\frac{(N^{\rm exp}_i-\alpha \, N^{\rm theo}_i )^2}
{ N^{\rm exp}_i}+ \frac{(\alpha -1)^2}{\sigma^2_{\rm syst}},
\label{concha_chi2}
\end{equation}
where  
$N^{\rm exp}_i$ and  $N^{\rm theo}_i$ are defined in the same way as in 
our $\chi^2$ definition in (\ref{def_chi2}). 
The $\chi^2$ was used, for example, in the KamLAND analysis in Ref.~\cite{concha04}.
We take, as in our analysis, $\sigma_{\rm syst}=4$\%. 
As we will see below, this choice of $\chi^2$ leads to even higher sensitivity 
of  $\theta_{12}$ measurement at SADO.
It is not surprising to obtain such result because the systematic uncertainty 
affects all bins simultaneously for $\chi^2$ in (\ref{concha_chi2}), 
whereas it can fluctuate bin-by-bin with our definition of $\chi^2$ in 
Eq. (\ref{def_chi2}).

Our second choice of $\chi^2$ is the one used for the analysis in 
reactor $\theta_{13}$ measurement \cite{munich}.
In such experiments we compare yields at identical 
near and far detectors and a large portion of the systematic errors cancels out 
in such a setting. Therefore, the statistical treatment involves the errors 
which have different characteristics, (un-)correlated between near and far 
detectors, and between bins.
Following \cite{munich,reactorCP}, we consider four types of systematic errors:
$\sigma_{\rm DB}$, $\sigma_{\rm Db}$, $\sigma_{\rm dB}$, and $\sigma_{\rm db}$.
The subscript D (d) represents the fact that the error is 
correlated (uncorrelated) between detectors.
The subscript B (b) represents that the error is correlated 
(uncorrelated) among bins.
The definition of $\chi^2$ is 
\begin{eqnarray}
 \chi^2_{\text{nf}}
 &\equiv&
 \min_{\text{$\alpha$'s}}
  \sum_{a = f, n}
          \left[
           \sum_{i=1}^{17}
            \left\{
             \frac{
              \left( N^{\rm theo}_{ai}
                 - ( 1 + \alpha_i + \alpha_a + \alpha ) N^{\rm exp}_{ai}
              \right)^2 }
              {   N^{\rm exp}_{ai}
                + \sigma_{\rm db}^2 (N^{\rm exp}_{ai})^2 }
             + \frac{ \alpha_i^2 }{ \sigma_{\rm Db}^2 }
            \right\}
           + \frac{ \alpha_a^2 }{ \sigma_{\rm dB}^2 }
          \right]
          + \frac{ \alpha^2 }{ \sigma_{\rm DB}^2 },
\label{NF_chi2}
\end{eqnarray}
where $N_{ai}^{\rm theo}$ represents the theoretical number of events
at the near ($a=n$) or far ($a=f$) detector within the $i$-th bin. 
Again, $N_{ai}^{\rm exp}$ is defined as the number of signal event 
calculated with the best-fit parameters of the ``experimental data''.  
See \cite{reactorCP} for more details. 

To simulate the reactor $\theta_{13}$ measurement with sensitivity 
up to $\simeq$ 1\% which is, very roughly speaking, equivalent to 
sensitivity up to $\sin^22\theta_{13} = 0.01$ 
for a long enough exposure (systematics dominated measurement),  
the following numbers have been taken for these errors \cite{reactorCP}; 
$\sigma_{\rm DB}=\sigma_{\rm Db}=2.5$\% and 
$\sigma_{\rm dB}=\sigma_{\rm db}=0.5$\%.
To simulate the systematic error of $\simeq$ 4\% at SADO we tentatively 
multiply by 4 all these errors. That is, we take 
$\sigma_{\rm DB}=\sigma_{\rm Db}=10$\% and 
$\sigma_{\rm dB}=\sigma_{\rm db}=2$\% 
in our analysis for SADO.
We also examine the optimistic case of systematic error of $\simeq$ 2\% at SADO, 
considering the possibility that the systematic error can be improved. 
The errors for this case are taken as 1/2 of the 4\% case. 
We have confirmed in the case of $\theta_{13}$ measurement that 
the error twice (four times) larger than that of \cite{reactorCP} leads to 
the sensitivity approximately equal to $\sin^22\theta_{13} = 0.02$ (0.04) 
for long enough exposure.

Since the front detector and SADO will be different in volume, assuming 
equal errors for both detector is nothing more than a crude approximation. 
But, we feel that it gives a reasonable framework to check the stability of 
our statistical treatment.

\begin{figure}[htbp]
\vglue -2cm
\begin{center}
\includegraphics[width=1.0\textwidth]{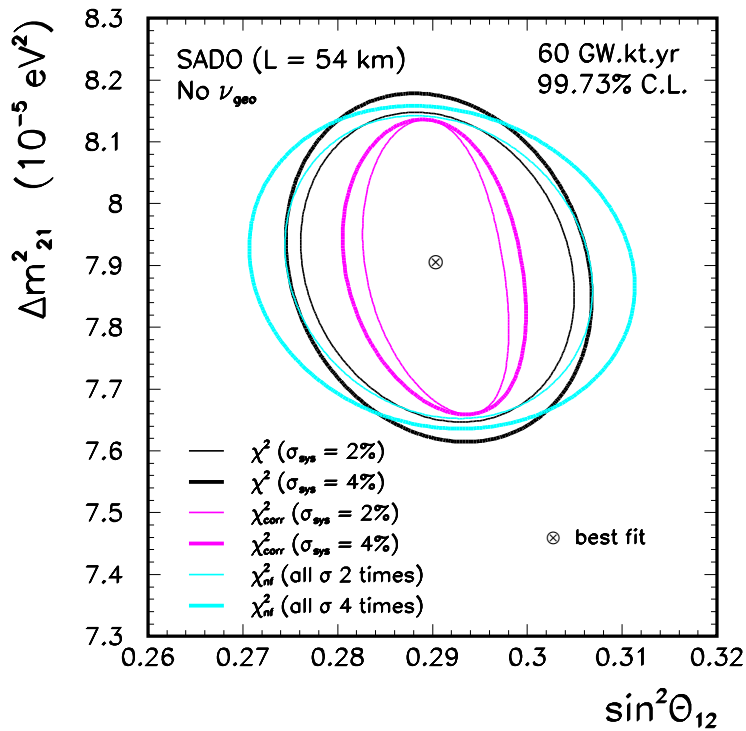}
\end{center}
\vglue -4cm
\caption{Regions allowed at 99.73\% CL for 60 $\GWth \cdot$kt$\cdot$yr 
exposure, determined by using three  different 
$\chi^2$ definitions: the definition used throughout this paper 
(Eq.~(\ref{def_chi2})), $\chi^2_{\rm corr}$ defined in Eq.~(\ref{concha_chi2})  
and $\chi^2_{\rm nf}$ defined in Eq.~(\ref{NF_chi2}).  
The contours with $\chi^2$ defined above are, in order, 
the middle (black line), the thinnest (red) and the thickest (blue) 
ones in widths. 
}
\label{chi2_test}
\end{figure}

In Fig.~\ref{chi2_test} we show the result of our analysis with three different 
definitions of $\chi^2$, (\ref{concha_chi2}) and (\ref{NF_chi2}) defined above, 
and the one given in (\ref{def_chi2}) that is adopted in our analysis. 
We observe that our result obtained in the 
previous section by using the $\chi^2$ in (\ref{def_chi2}) lies between 
the results of the analysis with two different $\chi^2$, in (\ref{concha_chi2}) 
and (\ref{NF_chi2}). It is expected that the result with $\chi^2$ in 
(\ref{concha_chi2}) gives tighter constraint on the oscillation parameters 
because the way how the systematic error is treated only allows fluctuation 
of the absolute normalization, not bin-by-bin independent fluctuations, 
and it is harder to mimic spectral shape distortion. 
We do not have any good physics intuition on what would be the result 
for the choice of $\chi^2$, in (\ref{NF_chi2}), but it turns out that 
the sensitivity is a little worse than our estimate with (\ref{def_chi2}).  
Nonetheless, the difference between the results obtained by using 
three different definitions of $\chi^2$ is not significant. 
Hence, we conclude that our estimate of the sensitivity of $\theta_{12}$ 
determination at SADO is stable under change of statistical treatment.

\section{Comparison of sensitivities of reactor $\theta_{12}$
measurement with other methods}
\label{comparison}

In this section, we compare the sensitivity calculated in Sec.~\ref{analysis}
for a reactor measurement of $\theta_{12}$ at SADO with the ones 
expected by other methods, in particular the solar neutrino 
experiments.
We assume CPT symmetry in the discussions in this subsection. 

To date the most accurate determination of $\theta_{12}$ can be
accomplished by combining 
all the solar neutrino experiments and KamLAND data~\cite{sno_salt}. 
According to Table 2 of Ref.~\cite{concha04},
the current solar neutrino data together with KamLAND result 
can be used to determine  $\sin^2{\theta_{12}}$ with about 
$\sim 8$\% precision at 68.27\% CL for 1 DOF. 
From the Table 3 of Ref.~\cite{bahcall-pena}, one expects that 
this precision will not be significantly reduced even if 
one combines with future KamLAND data corresponding
to 3 years of exposure. 

It is expected that if future solar neutrino experiments 
can detect $pp$ neutrinos selectively, they would give 
the best sensitivity to $\theta_{12}$ determination.  
Assuming the most optimistic 
error of 1\% for pp neutrino measurement and by combining with the other 
solar neutrino experiments as well as with 3 years running of KamLAND, 
one expects,  from Table 8 of Ref.~\cite{bahcall-pena}, 
that it is possible to measure  $\sin^2 \theta_{12}$ 
with uncertainty of $\sim 4$\% at 
68.27\% CL for 1 DOF. 
Despite the fact that this result is obtained with the previous 
best fit value of $\Delta m^2_{21}$ ($\sim 7\times 10^{-5}$ eV$^2$)  
it appears to be safe to assume that 
it will remain almost unchanged because the final precision of 
$\theta_{12}$ is essentially determined by the solar neutrino data.

The sensitivities expected by a reactor neutrino experiment calculated 
in the previous subsection should be compared with the precision 
obtained by all solar + KamLAND experiments 
mentioned above. This is done in Table~\ref{Tab3} from which 
one can see that SADO$_{\text{single,multi}}$ with exposure longer than  
20 $\GWth \cdot$kt$\cdot$yr, can determine $\theta_{12}$ better than
all other  observations combined. 
We want to mention that Gd-loaded Super-Kamiokande \cite{GdSK} 
cannot compete with the SADO's sensitivity on $\theta_{12}$ \cite{choubey}. 
Even after combining with the solar data the error 15\% on
$\sin^2 \theta_{12}$ at 99\% CL for 3 years operation 
is larger than SADO's 12\% at 99.97\% CL for its 
10 $\GWth \cdot$kt$\cdot$yr ($\simeq 0.5$ kt$\cdot$yr at SADO) exposure. 
It should be stressed that SADO alone can achieve such great precision 
without being combined with any other experiments, assuming 
that $\theta_{13}$ is measured with reasonable precision. 
Therefore, we conclude that a reactor measurement  
can supersede other methods for precise measurement of $\theta_{12}$ 
if the experimental systematic error of 4\% is realized.

\begin{table}
\caption[aaa]{
Comparisons of fractional errors of 
the experimentally determined mixing angle,  
$\delta s^2_{12}/s^2_{12} \equiv 
\delta (\sin^2\theta_{12}) / \sin^2\theta_{12}$,  
by current and future solar neutrino experiments and KamLAND (KL), 
obtained from Tables 3 and 8 of Ref.~\cite{bahcall-pena}, 
versus that by SADO$_{\text{single}}$ (SADO$_{\text{multi}}$)
obtained in this work at 68.27\% 
and 99.73\% CL for 1 DOF. 
}
\vglue 0.1cm
\begin{tabular}{c|cc}
\hline
\ Experiments\  & \ $\delta s^2_{12}/s^2_{12}$  at 68.27\% CL \ 
                & \ $\delta s^2_{12}/s^2_{12}$  at 99.73\% CL \  \\
\hline
 Solar+ KL (present)  & $ 8 $ \%  
                      & $ 26  $ \%   \\
\hline
 Solar+ KL (3 yr)  & $ 7 $ \% 
                   & $ 20 $ \%   \\
\hline
 Solar+ KL (3 yr) + pp (1\%) &  $ 4 $ \%  
                 & $ 11$ \%  \\
\hline
\multicolumn{3}{c}{54 km}\\
\hline
 SADO  for 10 $\GWth \cdot$kt$\cdot$yr  &  4.6 \% \   (5.0 \%)  
                 &  12.2 \%  \ (12.9 \%) \\
\hline
 SADO for 20 $\GWth \cdot$kt$\cdot$yr
                 &  3.4 \% \  (3.8 \%)
                 &  8.8 \% \  (9.5 \%) \\
\hline
 SADO for 60 $\GWth \cdot$kt$\cdot$yr
                 & 2.1 \%  \   (2.4 \%)
                 & 5.5  \% \   (6.2 \%) \\
\hline
\end{tabular}
\label{Tab3}
\vglue 0.5cm
\end{table}

Although our main concern in this paper is a precise determination 
of $\theta_{12}$ let us briefly discuss the sensitivity of SADO on 
$\Delta m^2_{21}$ and its comparison with the one by other methods. 
In Table~\ref{Tabmass} we present the expected sensitivity to 
$\Delta m^2_{21}$ at SADO$_{{\text{single}}}$ (SADO$_{{\text{multi}}}$)
at 1 $\sigma$ (68.27\%) and 3 $\sigma$ (99.73\%) CL for 1 DOF,
obtained with the prompt energy threshold of 0.9 MeV and with
the background effect  of geo-neutrinos from Fully Radiogenic Model.
The present KamLAND data together with all the solar neutrino data 
already achieved 14\% error at 3$\sigma$ CL on $\Delta m^2_{21}$ 
determination. 
As one can see from 
Table~\ref{Tabmass} (and Fig.~\ref{sado54b}), 
SADO$_{{\text{single}}}$ (SADO$_{{\text{multi}}}$)
will be able to reduce the error down to 
4.3\% (4.6\%) at 3$\sigma$ CL for 60 $\GWth \cdot$kt$\cdot$yr 
operation.
The error is smaller by a factor of 2 than that expected for 
the combined analysis of the future solar neutrino data and 
3 years operation of KamLAND found in Fig. 6 of Ref.\cite{bahcall-pena}. 
Nevertheless, it is still greater than the error of about 2.8\% at 
3$\sigma$ CL achievable by Gd-loaded Super-Kamiokande for its 
three years operation \cite{choubey}.

\begin{table}
\caption[aaa]{
The expected sensitivity to the $\Delta m^2_{21}$ at
SADO$_{{\text{single}}}$ (SADO$_{{\text{multi}}}$)
at 1 $\sigma$ (68.27\%) and 3 $\sigma$ (99.73\%) CL for 1 DOF,
obtained with the prompt energy threshold of 0.9 MeV and with
the background effect of  geo-neutrinos from
Fully Radiogenic Model.
}
\vglue 0.1cm
\begin{tabular}{c|cc}
\hline
\multicolumn{3}{c}{54 km}\\
\hline
\ exposure\  &\ \ $\Delta m^2_{12} \times 10^{5}$ eV$^2$ [68.27\% CL] \
             &\ \ $\Delta m^2_{12} \times 10^{5}$ eV$^2$ [99.73\% CL] \ \\
\hline
 10 $\GWth \cdot$kt$\cdot$yr
                 &\ \ $7.90^{+0.28}_{-0.30}$\ ($7.90^{+0.30}_{-0.30}$)
                 &\ \ $7.90^{+0.63}_{-0.83}$\
($7.90^{+0.70}_{-0.78}$)\\
\hline
 20 $\GWth \cdot$kt$\cdot$yr
                 &\ \ $7.90^{+0.21}_{-0.21}$\ ($7.90^{+0.21}_{-0.21}$)
                 &\ \ $7.90^{+0.48}_{-0.57}$\
($7.90^{+0.54}_{-0.56}$)\\
\hline
 60 $\GWth \cdot$kt$\cdot$yr
                 &\ \ $7.90^{+0.13}_{-0.14}$\ ($7.90^{+0.15}_{-0.14}$)
                 &\ \ $7.90^{+0.33}_{-0.35}$\
($7.90^{+0.36}_{-0.36}$)\\
\hline
\hline
\end{tabular}
\label{Tabmass}
\vglue 0.5cm
\end{table}


\section{Physics implications of accurate measurement of $\theta_{12}$}
\label{implication}

There exist a variety of physics implications available when 
accurate measurement of $\theta_{12}$ is carried out. 
Here, we describe only a part of them. 
Of course, $\theta_{12}$ is one of the fundamental parameters of 
particle physics and its precise determination is by itself clearly 
important, as already discussed in Introduction.
It may be appropriate to add a remark in this context. 
Namely, the determination of $\sin^2\theta_{12}$ to 2\% level 
is comparable in accuracy to that of the Cabibbo angle, 
which is about 1.4\% at this moment \cite{PDG}.
Therefore, SADO can open the new era in which we can enjoy 
balanced knowledge of lepton and quark mixings.

In the rest of this section, we focus  on points which may 
have greater impact on wider areas of research, including 
symmetry tests, solar astrophysics and the earth science. 
Let us start by discussing implications to solar astrophysics, 
and then move on to the other topics. 
We assume, apart from Sec.\ref{implication} D, the CPT invariance 
in our discussions.

\subsection{Observational solar astrophysics}

One of the purposes of the future solar neutrino experiments 
is to probe the deep interior of the Sun. We in fact live in an ideal 
location in the cosmos for this purpose, that is very close to the Sun, 
and the detailed information we can get for this one of the most 
standard main sequence stars should benefit wide area of astrophysics. 
In particular,  observation of the full spectrum of the solar neutrinos which 
extends from 10 keV to 10 MeV region must shed light on our 
understanding of stellar evolution.
It should provide us with the way of doing precise solar core 
diagnostics which is quite complementary to helioseismology.

It is expected that $pp$ neutrino flux will be calculated 
with error less than 2\% because of the 
powerful luminosity constraint. 
Now what is the most serious obstacle for doing stringent test 
for such an accurate prediction? 
How accurately we can measure the 
flux of $pp$ neutrinos which are about to leave the solar core? 
Assuming that the systematic error of the measurement can be controlled 
to less than 1-2\% level, the largest uncertainty comes from 
the error in $\sin^2\theta_{12}$ determination. 
Therefore, we need an independent measurement of $\theta_{12}$ 
to do precise observational solar physics.\footnote{
Even in a strategy by which $\theta_{12}$ can be determined simultaneously 
with the various components of solar neutrino flux by global fit \cite{bahcall-pena},  
precise information of $\theta_{12}$ should improve the accuracies of 
solar neutrino flux determination. 
}
Using the the most accurate value 
of $\theta_{12}$ which can be obtained by SADO, 
and using the observed solar neutrino flux at the Earth, 
we can determine accurately the infant solar neutrino flux 
at the solar core before they start to oscillate.

\subsection{Geo-neutrinos}
\label{geonu}

As described in Sec.~\ref{geo-sec}, the Earth is expected to be a very rich source of low 
energy ($E_\nu <3.4$ MeV) $\bar \nu_e$, whose detection can be of prime 
geophysical interest since they can provide otherwise inaccessible 
information on the abundance of radioactive isotopes such as U, Th and K
inside the Earth  and thereby help to unravel the internal structure and 
dynamics of the planet~\cite{geonu,Nunokawa:2003dd}.

It has been shown  that in a few years with a relatively small amount of 
exposure, KamLAND will have collected enough data in the energy  interval 
$0.9<E_{\text{prompt}}/{\text{MeV}}<2.6$ to clearly establish
the presence of geo-neutrinos~\cite{Nunokawa:2003dd}.
However, the precise determination of geo-neutrino flux 
requires considerably longer exposure 
(6 kt$\cdot$yr to determine U flux within 10\% uncertainty~\cite{Nunokawa:2003dd}).  
One of the reasons for the difficulty is that in this energy 
range, the neutrino flux from reactors is dominant at KamLAND 
and the geo-neutrino signal is predicted to be less than 1/5 of the 
reactor neutrino background. 
Therefore, precise measurement of 
$\theta_{12}$ is necessary to reliably subtract  the background 
contribution from reactors in order to determine precisely 
the flux of geo-neutrinos.

The high-precision determination of $\theta_{12}$ by SADO 
explored in this paper is essentially unaffected by the geo-neutrino 
background, as shown in Sec.\ref{analysis} (See Fig.~\ref{sado54b}). 
Then, we can use this information to subtract the dominant
contribution from reactors for the energy 
range relevant for geo-neutrinos, 
$0.9<E_{\text{prompt}}/{\text{MeV}}<2.6$,  
at KamLAND. 
We note that SADO itself would not be adequate to measure the 
geo-neutrino flux, because the geo-neutrino flux contribution 
in the energy interval
$0.9<E_{\text{prompt}}/{\text{MeV}}<2.6$ is expected 
to be at most $\simeq$ 10\% at SADO. It is due to the fact that 
it is so close to the Kashiwazaki-Kariwa NPP, from which it receives 
a dominant contribution of reactor antineutrinos.

\subsection{Determination of masses and mixing parameters in future 
neutrino experiments}

The precise measurement of $\theta_{12}$ should help identify the 
unknown quantities and to improve the accuracy of determination 
in future neutrino experiments. 
We mention here only two of them, neutrinoless double beta decay and 
CP phase measurement in LBL experiments. 
Neutrinoless double beta decay is probably the most promising way 
of identifying the absolute scale of neutrino mass in laboratory experiments 
\cite{review2}.
As is well known the constraint imposed by the experiments on either the 
mass scale or the other observables such as Majorana phases involves 
$\theta_{12}$. See e.g., \cite{dbeta} and the references cited therein. 
Its precise knowledge, therefore, should help identifying these quantities.

Discovery of Leptonic CP violation in LBL experiments is one of the 
challenging goals in future neutrino experiments. 
The CP violating piece of the appearance oscillation probabilities 
$P(\nu_\mu \to \nu_e)$ and $P(\bar \nu_\mu \to \bar\nu_e)$ is proportional 
to $\Delta m^2_{21}$ and $\sin{2\theta_{12}}$. 
Since detecting CP violating effects requires enormous precision 
in the experiments, uncertainties in these relevant parameters 
could easily obscure the discovery of CP violation. 
Moreover, once detection of CP violation is done, it will motivate and 
facilitate actual determination of the value of the CP angle $\delta$. 
It does require very precise determination of 
$\Delta m^2_{21}$ and $\sin^2{\theta_{12}}$ as well as 
$\Delta m^2_{31}$ and $\sin^2{\theta_{23}}$, though the latter 
are not the subject in this paper. 
Therefore, accurate measurement of these quantities is the prerequisite 
for determination of $\delta$.

\subsection{Test of CPT symmetry}

CPT symmetry is one of the most fundamental symmetries which 
are respected in relativistic quantum field theory. 
It implies that the masses and the mixing angles of particles 
and their antiparticles are identical. Then, one can perform a 
stringent test of CPT symmetry in lepton sector by accurately 
measuring $\theta_{12}$ and $\Delta m^2_{21}$ with use of reactor 
antineutrinos and by comparing them with the same parameters 
measured by solar neutrino experiments.

In the previous section we have shown that the sensitivity 
attainable in reactor measurement of $\theta_{12}$ improves, 
in a reasonable setting, the one achievable by future solar 
neutrino experiments. Then, the significance of the CPT test 
is controlled by the accuracy reached by solar neutrino measurement. 
By comparing our predictions with those presented in 
Ref.~\cite{bahcall-pena} we can estimate the future sensitivity to 
CPT test in this sector.
We note that the sensitivity on $\theta_{12}$ in \cite{bahcall-pena}
essentially comes from solar neutrino data, not from KamLAND, 
as the latter is more efficient in decreasing the range of 
$\Delta m^2_{12}$ but not of $\theta_{12}$.  

The current bound on the difference between
$\sin^2{\theta_{12}}$ for neutrino and
$\sin^2{\bar{\theta}_{12}}$ for antineutrinos
is rather weak \cite{KamLAND_new}.
Even if we assume that $\bar{\theta}_{12}$ is in the first 
octant,\footnote{
If we allow $\bar{\theta}_{12}$ to be in the entire quadrant,
$0 \leq \bar{\theta}_{12} \leq \pi/2$, an another solution
$\bar{\theta}_{12} ^{\prime} = \pi/2 - \bar{\theta}_{12} $
emerges \cite{carlos04}, and SADO's measurement will not be able to
improve the CPT bound.
}
\begin{eqnarray}
|\sin^2{\theta_{12}} - \sin^2{\bar{\theta}_{12}}| \lsim 0.3,
\label{CPT_theta12curr}
\end{eqnarray}
at 99.73\% CL, which can be improved by a factor 5
in the future if the accuracy expected from future solar neutrino
data~\cite{bahcall-pena} is achieved and SADO is realized.
See Table~\ref{Tab3}.

For mass squared differences, the current bound~\cite{carlos04} is 
\begin{equation} 
|\Delta m^2_{21} - \bar{\Delta m^2_{21}} |\leq 1.1 \times 10^{-4} \text{eV}^2
\end{equation} 
at 99.73\% CL, where $\bar{\Delta m^2_{21}}$ is the mass squared 
difference of antineutrinos.
We observe that SADO will not be able to make significant 
improvement of this bound beyond what can be reached by combining 
future solar with KamLAND data since the bound will be essentially
determined by the uncertainty of $\Delta m^2_{12}$ coming
from solar neutrino data which is significantly larger than 
that of $\bar{\Delta m^2_{21}}$ determined by KamLAND or SADO.

For comparison, let us mention that the only available constraint 
on $\theta_{23}$ is obtained by Gonzalez-Garcia  {\it et al.} 
\cite{concha_CPT} by analyzing atmospheric neutrino data. 
The bound obtained for the difference between neutrino and 
antineutrino mixing angles is:
\begin{eqnarray}
- 0.41 \leq 
\sin^2{\theta_{23}} - \sin^2{\bar{\theta}_{23}} \leq 
0.45,
\label{CPT_theta23}
\end{eqnarray} 
at 99.73\% CL level.

For $\Delta m^2_{32}$ there are two results on CPT test, one by 
Super-Kamiokande group \cite{saji_noon04}
\begin{eqnarray}
- 1.9 \times 10^{-2} \mbox{ eV}^2 \leq 
|\Delta m^2_{32}| - |\bar{\Delta m^2_{32}}| \leq 
4.8 \times 10^{-3} \mbox{ eV}^2, 
\label{CPT_SK}
\end{eqnarray} 
and the other by Gonzalez-Garcia {\it et al.} \cite{concha_CPT}
\begin{eqnarray}
- 10^{-2}  \mbox{ eV}^2 \leq 
\Delta m^2_{32} - \bar{\Delta m^2_{32}} \leq 
3.4 \times 10^{-3} \mbox{ eV}^2, 
\label{CPT_Dm2}
\end{eqnarray} 
both at 99.73\% CL level~\footnote{
The bound by Super-Kamiokande was obtained by using the
two-flavor analysis of atmospheric neutrino data, so that we have
interpreted it to be the one placed on absolute values of $\Delta
m^2_{32}$.
The bounds obtained in Ref.~\cite{concha_CPT} are based on
the analysis assuming the normal mass ordering for both neutrinos
and antineutrinos.
But, since both solar and KamLAND data imply that the splittings are
hierarchical for both neutrinos and antineutrinos, and the 13 mixing
angles for neutrinos and antineutrinos are favoured to be small,
the bound is not expected to be quantitatively very different if
neutrinos and/or antineutrinos masses have inverted ordering.}.

\subsection{Test of Non-standard neutrino interactions}

Precise measurement of $\Delta m^2_{21}$ and $\theta_{12}$ 
by reactor experiments have an important impact 
in constraining the non-standard neutrino 
interactions (NSNI) with matter~\cite{NSNI}. 
As discussed recently in ~\cite{NSNI_recent}, 
the currently allowed parameter region by solar and KamLAND 
data could be significantly modified were the NSNI present. 
It can be tested to a better accuracy by future data from KamLAND. 
The possible presence of such NSNI can be further constrained 
(or confirmed) by performing the precise measurement of $\Delta m^2_{21}$ 
and $\theta_{12}$ with a SADO type reactor neutrino experiment.
The point is that while solar neutrinos are severely affected by 
the matter effect reactor neutrinos are not. 
Therefore, such NSNI can be strongly constrained, 
assuming the CPT symmetry, if the mixing parameters inferred 
from solar and reactor neutrino observations coincide with each other.

\section{Conclusion}

In this paper, we have investigated the potential power of dedicated 
reactor neutrino experiments for precision measurement of $\theta_{12}$. 
By placing a detector in an appropriate baseline distance from 
a powerful nuclear reactor complex, and assuming 4\% systematic error, 
a world-record sensitivity on  
$\sin^2{\theta_{12}}$  $\simeq $ 2\% ($\simeq$ 3\%) at 
68.27\% CL is shown to be attainable 
by 60 GW$_{\text{th}}\cdot$kt$\cdot$yr 
(20 GW$_{\text{th}}\cdot$kt$\cdot$yr) operation, superseding 
all the other proposed methods. 
Thus, it improves, after 20 $\GWth \cdot$kt$\cdot$yr operation  
the accuracy to be achieved jointly  by KamLAND and the 
existing solar neutrino experiments for 3 years by more than a factor of 2. 
At 60 $\GWth \cdot$kt$\cdot$yr operation its enormous sensitivity of 
approximately 2\% is about a factor of 2 better than that to be reached 
by additional $^7$Be and $pp$ neutrino observation with extremely small 
total errors of 5\% and 1\%, respectively (see Table~\ref{Tab3}).

Toward the conclusion, we have carried out a careful estimation of the 
optimal baseline distance and obtained a rather wide range, between 
50 and 70 km, from the reactor neutrino source for the current 
best fit value of $\Delta m^2_{21}$. 
The distance is determined by the requirements that the first oscillation 
maximum occurs around a peak energy of the event number distribution in the 
absence of oscillation, 
and the geo-neutrino background is harmless.
We have checked by taking a detector placed at 54 km from a reactor 
neutrino source that geo-neutrino background does indeed not produce 
any significant additional errors.

To estimate the effect of background caused by other surrounding 
reactors we have examined a concrete setting. 
We took the Kashiwazaki-Kariwa NPP in Japan, the most powerful 
reactor complex in the world as a reference,  
and assumed a detector (SADO) located in Mt. Komagatake 
which is 54 km away from the reactor complex. 
We have verified that the uncertainty due to the other 15 NPPs 
produce only 20\% increase in the error of $\sin^2{\theta_{12}}$ 
determination (See Tab.~\ref{Tab2}).
We emphasize that since nuclear reactors are less populated in most of 
the rest of the world, examination of SADO with 15 NPPs will give us a 
conservative estimate of sensitivities for the similar reactor $\theta_{12}$
experiments on the globe.

Of course, $\theta_{12}$ is one of the fundamental parameters of 
particle physics and the significance of its precise determination  
by itself cannot be overemphasized. Nonetheless, we have discussed 
that there is  a plethora of interesting physics implications 
available when such an accurate measurement of $\theta_{12}$ is carried out.
We point out its impact to solar astrophysics, geophysics, determination 
of the yet unknown mixing parameters, CPT symmetry test, and exploring 
non-standard interactions.


\newpage

\appendix

\section{Thermal powers and the distances to the detectors} 
\label{ap1}
In Table~\ref{Tabzero} we present the 
thermal powers of the 16 NPPs in Japan and their distances 
to KamLAND and SADO at Mt. Komagatake which are used in our analysis.

\begin{table}[th]
\caption[aaa]{
The maximal thermal powers of the 16 NPPs in Japan and their distances 
to KamLAND and SADO at Mt. Komagatake are presented in units of 
GW$_{\text{th}}$ and of km, respectively. 
Their fractional contributions of $\bar{\nu}_{e}$ flux without 
oscillation to each detector, denoted as $F_{\text{KL,SADO}}$, are also shown in \%.  
The thermal powers of current operation are used for KamLAND analysis and 
the numbers and the distances are taken from \cite{tese}. 
For SADO sensitivity analysis, the future values for thermal powers are adopted for 
Hamaoka and Shika (indicated in parentheses), and the values are taken 
from the following web sites:\\
http://www.chuden.co.jp/hamaoka/DETAIL/newgo-setsubi.html\\
http://www.rikuden.co.jp/shika/outline2/\\
$L_{\text{SADO}}$ are calculated by using 1/25000 map, 
assuming that the Earth is a perfect sphere with radius of 6380 km.
}
\vglue 0.1cm
\begin{tabular}{c|c|cc|cc}
\hline
          &Thermal Power (GW$_{\text{th}}$)   & \   $L_{\text{KL}}$ (km) 
& \  $F_{\text{KL}}$ (\%) & \ $L_{\text{SADO}}$ (km) & \  $F_{\text{SADO}}$ (\%) \ \  \\
NPP   &    &   & & &      \\
\hline
Kashiwazaki   &    24.3   &    160 &  32.0  &    54  &  77.0      \\
Ohi           &    13.7   &    179 &  14.4  &    354 &  1.0       \\
Takahama      &    10.2   &    191 &  9.4   &    348 &  0.8      \\
Hamaoka (future)   & 10.6 (14.5)   &  214 &  7.8   &    300 & 1.5       \\
Tsuruga       &    4.5      &    138 & 7.8    &    299 & 0.5       \\
Shika  (future) &   1.6 (5.5) &    88  & 8.2   &    158 & 2.0       \\
Mihama        &    4.9    &   146  &  7.8   &   307 & 0.5       \\
Fukushima-1   &  14.2       &   349  & 3.9    & 144 & 6.3       \\
Fukushima-2   &  13.2     &   345  &  3.7   & 139 &  6.3       \\
Tokai-II      &    3.3    &   295  & 1.3    &  109 & 2.6       \\
Shimane       &   3.8     &   401  & 0.8    &  492 & 0.1       \\
Ikata         &    6.0    &   561  &  0.6   &  697 & 0.1       \\
Genkai        &   10.1    &   755  & 0.6    &  828 & 0.1       \\
Onagawa       &    6.5    &   431  & 1.2    &    236 &  1.07      \\
Tomari        &    3.3      &   783  & 0.2    &  606 & 0.08      \\
Sendai        &    5.3    &    830 & 0.3   &  953 &  0.05      \\
\hline
\end{tabular}
\label{Tabzero}
\vglue 0.5cm
\end{table}

\newpage 

\section{Details of Analysis Procedure} 
\label{ap2}
Here we describe some details of our analysis procedure. 
We compute the expected number of $\bar \nu_e$ events in the 
$i$-th energy bin, $N_i^{\text{theo}}= N_i^{\text{reac}} + N_i^{\text{geo}}$, 
where $N_i^{\text{reac}}$ and $N_i^{\text{geo}}$ are computed as follows
\begin{eqnarray}
N_i^{\text{reac}}(\sin^2 \theta_{12},\Delta m^2_{21}) &=&
N_p T 
\nonumber \\
\times \int &dE_\nu& \, \sum_{k=1}^{N_{\text{\tiny NPP}}} \epsilon_k
\phi_k(E_\nu) P(\bar{\nu}_e \to \bar{\nu}_e; L_k, E_\nu) 
\sigma(E_\nu) \int_i dE 
\epsilon_{\text{det}}\,R(E,E^\prime),
\end{eqnarray} 
and 
\begin{eqnarray}
N_i^{\text{geo}}(\sin^2 \theta_{12},\phi_{\text{U}}) &=&
N_p T \, (1-\displaystyle \frac{1}{2} \sin^2 2\theta_{12}) 
\nonumber \\
&\times&\int dE_\nu \, 1.83\, \phi_{\text{U}}\, [f_{\text{U}}(E_\nu)
+ f_{\text{Th}}(E_\nu)] \sigma(E_\nu) \int_i dE 
\epsilon_{\text{det}}\,R(E,E^\prime),
\end{eqnarray} 
where $N_p$ is the number of target protons in the detector 
fiducial volume,  $T$ is the exposure time, and 
$\phi_k(E_\nu)$  is the neutrino flux spectrum from 
the $k$-th NPP expected at its maximal thermal power operation.
$\epsilon_k$ denotes the averaged operation efficiency of 
the $k$-th NPP for a given exposure period and it is 
taken to be 100\% here under the understanding that the unit we use 
GW$_{\text{th}}\cdot$kt$\cdot$yr refers the actual thermal power 
generated, not the maximal value. 
$P_k(\bar{\nu}_e\ \to \bar{\nu}_e, L_k, E_\nu)$ is the familiar 
antineutrino survival probability in vacuum 
(given by Eq.(\ref{prob}) with $\theta_{13}=0$ and by setting $f_i=\delta_{ik}$) 
for the $k$-th NPP,  
and it explicitly depends on $\Delta m^2_{21}$ and  $\sin^2 \theta_{12}$.
$\sigma(E_\nu)$ is the $\bar{\nu}_e$ absorption cross-section 
on proton, $\epsilon_{\text{det}}=0.898$ is the detector efficiency and 
$R(E,E')$ is the energy resolution 
function, which is assumed to have Gaussian form, 
$E$ ($=E_{\text{prompt}})$ the observed prompt energy (total $e^+$ energy) 
and $E'=E_{\nu}$ - 0.8 MeV 
the true one.  

We assume, following KamLAND \cite{KamLAND_new}, the energy 
resolution to be $\sigma(E)/E=6.2\%/\sqrt{E(\text{MeV})}$ and consider each 
energy bin to be 0.425 MeV wide. The summation over $k$ is meant to 
sum over the contributions from all the reactor NPPs, 
{\em i.e.}, $N_{\text{\tiny NPP}}=1$ or $16$ 
depending upon whether only the Kashiwazaki-Kariwa or all 16 reactors are 
considered. We have used in our calculations $E_{\text{prompt}}>0.9$ MeV 
and the future thermal powers of Hamaoka and Shika (see Table~\ref{Tabzero}). 
Other relevant informations can be found in Ref.~\cite{tese}.

We have performed  the geo-neutrino calculation as in Ref.~\cite{geonu}.   
$\phi_{\text{U}}$ is the total flux of geo-neutrinos from U decays,  
we assume that the total flux of geo-neutrinos from Th 
decay is 83\% of  $\phi_{\text{U}}$, $f_{\text{U}}$ and $f_{\text{Th}}$ 
are the normalized energy distributions of U and Th geo-neutrinos.
Here we have assumed the oscillation probability to be averaged. 
The procedure is not exactly correct but for our current purposes 
it gives a good approximation.

\newpage

\begin{acknowledgments}
We thank Fumihiko Suekane for triggering our interests in 
the reactor measurement of $\theta_{12}$ and for numerous correspondences. 
We also acknowledge useful correspondences with 
Carlos Pe{\~n}a-Garay on the determination of the oscillation parameters 
by solar neutrinos, and with M. C. Gonzalez-Garcia, 
Andr{\'e} de Gouvea, Choji Saji, and Masato Shiozawa 
on tests of CPT symmetry. 
Thomas Schwetz kindly informed us of Ref.~\cite{bouchiat}.
H.M. is grateful to PUC, Rio de Janeiro, for enjoyable visit 
during which the first draft of this paper was written. 
Three of us (H.M., H.N. and R.Z.F) acknowledge 
the theory group of Femilab where 
the final part of this work was done. 
This work was supported by the Grant-in-Aid for Scientific Research
in Priority Areas No. 12047222, Japan Ministry
of Education, Culture, Sports, Science, and Technology, 
the Grant-in-Aid for Scientific Research, No. 16340078, 
Japan Society for the Promotion of Science and by 
Funda{\c c}{\~a}o de Amparo {\`a} Pesquisa do Estado de S{\~a}o 
Paulo (FAPESP) and Conselho Nacional de  Ci{\^e}ncia e Tecnologia (CNPq).
\end{acknowledgments}


\end{document}